\documentclass[11pt]{article} 

\usepackage{amsmath,amsthm,latexsym,amssymb,amsfonts,epsfig}

\newcommand{\be}{\begin{equation}}
\newcommand{\ee}{\end{equation}}
\newcommand{\ba}{\begin{eqnarray}}
\newcommand{\ea}{\end{eqnarray}}

\title{{\sf Scalar Material Reference Systems}\\
{\sf and Loop Quantum Gravity}} 
\author{
{\sf K. Giesel}$^1$\thanks{{\sf 
kristina.giesel@gravity.fau.de}},
{\sf T. Thiemann}$^1$\thanks{{\sf 
thomas.thiemann@gravity.fau.de}}\\ 
\\
{\sf $^1$ Institute for Quantum Gravity (IQG)} \\ {FAU Erlangen -- N\"urnberg,}\\
{\sf Staudtstr. 7, 91058 Erlangen, Germany}\\
}
\date{{\small\sf \today}}

\makeatletter
\@addtoreset{equation}{section}
\makeatother

\begin{document} 

\maketitle

{\sf
\begin{abstract}
In the past, the possibility to employ (scalar) material reference systems 
in order 
to describe classical and quantum gravity directly in terms of gauge invariant
(Dirac) observables has been emphasised frequently. This idea has been 
picked up more recently in Loop Quantum Gravity (LQG) with the aim to 
perform a reduced phase space quantisation of the theory thus possibly 
avoiding
problems with the (Dirac) operator constraint quantisation method for 
constrained system.

In this work, we review the models that have been studied on the classical
and/or the quantum level and parametrise the space of theories so far 
considered. We then describe the quantum theory of a model that, to the best
of our knowledge, so far has only been considered classically. This model
could arguably called the optimal one in this class of models considered as it 
displays the simplest possible true Hamiltonian while at the same time 
reducing all constraints of General Relativity.
\end{abstract}
}

\newpage

\section{Introduction}
\label{s1}

In series of seminal papers
\cite{0,1,2,3}, Kucha\v{r} and his collaborators have 
constructed a whole class of manifestly spacetime diffeomorphism invariant
matter actions which have the remarkable feature to clarify the conceptual
setup of General Relativity. A further model of this type, originally introduced by Rovelli and Smolin in \cite{4a}, was analysed by Kucha\v{r} and collaborators in \cite{4}. The motivation for introducing those matter 
actions is to use the matter considered as 
a material reference system. This allows 
  to isolate the pure gauge degrees of freedom (under
spacetime diffeomorphisms) encoded in general relativity in an elegant way and yields to theories where gauge invariant (Dirac) 
observables can be constructed by means of those matter reference fields. The matter considered in 
\cite{0,1,2}
was coined dust matter because its energy momentum tensor is that of a 
pressure free perfect fluid. The advantage of the availability of 
such a description of the phase space of matter and geometry with regard 
to quantum gravity was already emphasised in those works although a concrete
quantum framework was not available at that time and the quantum theories following from this models could only be discussed at a formal level. 

A particular feature of the matter models considered is that they lead to a deparametrised form of general relativity (plus, if considered, standard model matter) and that the resulting hypersurface deformation algebra for the Hamiltonian constraints, that is the algebra among the individual constraints, becomes Abelian. This can be achieved by writing those Hamiltonian constraints  in an equivalent\footnote{I.e. they define
the same constraint surface and the same Dirac observables.} form 
in which they are linear in the reference matter momenta. This is an important
property because in this form the constraints form a true Lie algebra in contrast to the Hamiltonian constraints in its standard form. As far as the quantum theory is considered, this is an advantage because many techniques that aim at solving the constraints
in the quantum theory (such as group averaging \cite{5}) cannot be applied when there occur structure functions instead of structure constants. 

The equivalent form of the Hamiltonian constraint that has mutually 
commuting Poisson brackets with itself is of the typical form 
$C^{{\rm tot}}(x)=P(x)+h(x)$ 
where $P$ is the momentum conjugate to one of the dust fields $T$ (called the 
clock field) and $h$ is a scalar density of weight one built from the 
non dust, spatial scalar density weight one contribution $C$ to the 
Hamiltonian constraint, the 
(square root of) the determinant
$Q=\sqrt{\det(q)}$ of the spatial metric $q_{ab}$, the density two 
spatial scalar $D=q^{ab} C_a C_b$ where $C_a$ is the non dust contribution 
to the spatial diffeomorphism constraint, the density one respectively 
zero spatial 
scalars $V=q^{ab} T_{,a} C_b, \; U=q^{ab} T_{,a} T_{,b}$. 
For a certain subclass of models, the dependence
of $h$ on $V,U$ is trivial and it then follows from the first class property
of the constraint algebra that the $h(x)$ themselves must be mutually 
vanishing. 

In \cite{6} the general solution of phase space functions $h(x)$ of density 
weight $w$ built from $C(x),Q(x),D(x)$ (but not on $V(x), U(x)$) with mutually
commuting Poisson brackets was found in terms of a first order PDE which 
can be solved by the method of characteristics, thus 
extending the set of solutions 
that were obtained from the concrete models \cite{0,1,2,4a,4} but without 
providing concrete covariant actions from which they result. In \cite{7} it was
shown that all of these solutions can be obtained by using covariant actions 
of the Lagrange multiplier type used in \cite{0,1,2,3,4a,4}.

On the other hand, one can ask whether there are covariant scalar matter models 
that do not rely on Lagrange multipliers at all. This question was studied 
in \cite{8} and it was pointed out that provided that the scalar field has 
pure derivative coupling to the gravitational field then such models 
again give rise to mutually Poisson commuting Hamiltonian densities $h$. 
Also the inverse question can be analysed using the methods developed in 
\cite{8}, namely, given a general solution $h$ as described in \cite{6} 
(with density weight one), what is the Lagrangian $L$ that gives rise to it?
The solution, when it exists, can be written implicitly as the solution 
of an ODE. 

The general framework for describing the reduced phase space of a constrained
system with a generally covariant Lagrangian in manifestly gauge invariant 
form has been considered, to the 
best of our knowledge, for the first time in \cite{9,10}. Aspects of this 
have been rediscovered in different language by several authors, e.g. 
\cite{11,12,13}. In these works, a complicated formula appears that ``projects''
a non gauge invariant function $f$ on phase space to a (weakly) gauge invariant 
one. It involves an infinite series of multiple Poisson brackets between 
$f$ and the Hamiltonian constraint which makes it apparently impractical to 
use. However, as pointed out in \cite{14}, the Poisson algebra of these formal 
observables can nevertheless still be simple, thus enabling in principle a 
concrete reduced phase space quantisation approach to quantum gravity.

In \cite{15,16} the frameworks of \cite{0,9,10} were combined in order to 
compute the classical reduced phase space of \cite{0}. Namely, in 
\cite{0,1,2,4a,4} the reduced phase spaces where only described for some models in suitable gauges and the last missing step before arriving at the fully reduced 
phase space was to use the projector formula provided in \cite{9,10} and construct general observables with respect to the Hamiltonian constraint. 

 The reduced phase space described in \cite{15,16} was then quantised using the technology of Loop 
Quantum Gravity \cite{17,18} in \cite{19} and thus providing for the first 
time a concrete model for LQG where {\it all constraints are solved}. This means that 
all operators in question are {\it Dirac observables} and time evolution 
is driven by a {\it physical (i.e. non vanishing, gauge invariant) 
Hamiltonian} for which a concrete quantisation was provided. 
In \cite{19} two possible quantisations of the reduced physical phase space 
were discussed. In the first one the usual
kinematical Hilbert space of LQG becomes the {\it physical Hilbert space}, 
no solutions of the constraints have to be computed, the usual LQG
inner product is the {\it physical inner product} here.
In the second possibility techniques from the Algebraic Quantum Gravity 
framework introduced in \cite{23} are considered and the quantum theory 
is defined on an abstract algebraic graph using von Neumann's infinite 
tensor product Hilbert space.
\\
\\
Later, in \cite{20}, a scalar matter model of the type considered in \cite{8},
namely the model of \cite{4a,4}
was used in order to perform the reduced phase space quantisation by methods
from LQG of General
Relativity with a matter content different from \cite{19}. 
While in \cite{19} four scalar reference fields are considered the model in \cite{20} contains only one scalar field.
The most 
significant difference between these works is that the spatial diffeomorphism
constraint is solved already classically in \cite{19} using three out of the four reference fields while in \cite{20}
it needs to be solved at the quantum level.
\\
\\
A reduction of the diffeomorphism constraint at the classical level 
has the advantage that one can avoid ambiguities that can occur in the quantum reduction. These exists also for the spatial diffeomorphism constraint and are in general not unproblematic because those could possibly present an anomaly of the algebra of observables
\cite{21}. It turns out that the classical reduction with respect 
to the spatial diffeomorphism constraints of the dust model of
\cite{0} has the consequence, when using the standard LQG representation, 
that the physical Hamiltonian of the theory (a diffeomorphism invariant 
quantity) needs to be implemented as an operator that 
preserves the underlying graph on which the quantum states are 
defined\footnote{This simply follows from the result that spatially 
diffeomorphism invariant operators that are graph-modifying are 
not densely defined in the standard kinematical representation of LQG 
as shown in in \cite{27}}. 
If one wants to avoid a graph-preserving quantisation which was one of 
the motivations for the quantum model introduced in \cite{20}, then 
within LQG one is 
forced to solve the spatial diffeomorphism constraint at the quantum 
level as shown and implemented for the first time in \cite{32}. In this case 
the operators involved are not knot class preserving.\\
\\
As far as semiclassical aspects of the models are concerned, there are two
issues to be keppt in mind. First of all, semiclassical tools have not been
developed at the level of the Hilbert space of spatially diffeomorphism
invariant states at all, neither in the knot class preserving case nor 
in the non preserving case. Secondly, at the kinematical level, only for 
graph-preserving quantisations do exist 
semiclassical tools. These have been used to establish the correctness of 
the semiclassical limit of the dynamics for short time scales 
in \cite{19} using techniques developed in \cite{23}. 
However, for graph-modifying dynamical operators that occur 
the model in \cite{20} one first needs to develop semiclassical states 
first for diffeomorphism invariant states to begin with and then in addition
for knot class modifying operators 
before the semiclassical sector of the dynamics can be investigated.
Notice that on the other hand in the Algebraic Quantum Gravity (AQG) framework
of \cite{23} here is no need to preserve algebraic subgraphs on which the 
algebraic spin network functions depend.
\\
\\
Another more recent model discussed in \cite{22} that was also already described in 
 the appendix of \cite{0} as a special case of the model in \cite{0} results 
by switching off three Lagrange multiplier fields in the general Lagrangian
of \cite{0} which leads to a model with only one Lagrange reference matter field and likewise to \cite{20}
 deparametrises only the Hamiltonian constraint but does not rewrite the spatial diffeomorphism constraints in equivalent form that they are linear in the momenta of the reference fields as in \cite{19}.
\\
 The authors nevertheless claim that they 
can treat their model with the methods from \cite{23} without having 
to care about 
the spatial diffeomorphism constraint anymore and nevertheless describe the physical sector
of the theory. Moreover, they claim that 
they have found the most promising classical foundation
for the reduced phase space quantisation of General Relativity using
methods from LQG in the sense that the physical Hamiltonian simplifies 
in the maximal sense. As we will see, some of these claims are too strong.
\\
\\
It transpires that there are several models of General Relativity, the 
standard model matter and certain additional matter in the literature 
which lead to different reduced phase spaces and different physical 
Hamiltonians with different properties. It would therefore be desirable 
to start from a common platform in the form an action principle, to 
study the different models from a unified viewpoint and to analyse their 
different physical properties and promises. This is the objective of the 
present paper.\\   
\\
This paper is architectured as follows:\\
\\
In section two we review the known results from the literature 
and then construct a general Lagrangian that encompasses and extends all 
models considered so far in the literature with minimal coupling
of the metric (no derivatives) to matter. It serves as the desired 
common platform or parametrisation of theory space. Different models can be 
reached by switching on and off certain parameters. Basically the models 
considered so far fall into two types: I. Those which just deparametrise 
time and II. those which deparametrise all of spacetime. We describe their 
Hamiltonian analysis in a unified fashion.
\\
\\
In section three we reconsider the model of type II studied in 
\cite{22} and iterate arguments well known in the literature 
(e.g. \cite{24,25} and references therein) that reveal that 
some of the claims made in \cite{22} are not tenable. 
It is is not possible to simply drop the spatial diffeomorphism constraint and quantising using the algebraic quantum gravity framework as claimed by the authors.
Rather one has to treat the quantisation of the dynamics in that model more carefully. A possible way to deal with the spatial diffeomorphism constraint in the algebraic framework is to quantise it by the Master constraint method, see  \cite{24} for a discussion, 
if one wants to proceed along \cite{23}. In \cite{20} the reduction of the spatial diffeomorphism constraints at the quantum level was performed via the refined algebraic quantisation technique \cite{5} by means of a rigging map yielding the usual (spatially) diffeomorphism invariant Hilbert space of LQG. The physical Hamiltonian was then quantised on the diffeomorphism invariant Hilbert space.
\\
\\ 
Moreover, a careful treatment shows that there is an important issue  
with the choice of sign of a square root that has not been mentioned in 
\cite{22}. Namely, if 
one insists on the usual energy conditions for the energy momentum tensor,
then the sign of the physical Hamiltonian must be constrained in that model
whence the physical Hamiltonian is still not without a square root as claimed,
since it is constrained to equal plus or minus its absolute value. 
We also show that
the model in \cite{22} corresponds to a subsector of the model \cite{19}
which also explains why there is still an absolute value (and thus a square
root) involved in the physical Hamiltonian of \cite{22}. The fact that 
this is a subsector of \cite{19} corresponding to vanishing spatial 
diffeomorphism constraint makes it also transparent that the latter 
cannot be dropped.\\
\\
In section four we work out the details of the LQG quantisation of the 
model \cite{2} following the procedure of \cite{15,16,19}. This model could 
arguably called the optimal one in the class of models considered so far
in the sense that not only also the spatial diffeomorphism constraint 
is already reduced classically, thus avoiding the ambiguities and potential 
problems pointed out in 
\cite{21} in connection with the scalar product on spatially diffeomorphism 
invariant states, but furthermore, this time there is really no square root 
involved in the 
physical Hamiltonian.
\\
In section five we summarise and conclude.

\section{Theory Space}
\label{s2}

There are roughly two types of scalar field models that have been studied
in the literature and that were found convenient for investigations in quantum
gravity. Models of type I are based on Lagrangians that involve beside the fields that 
play the role of the reference fields additional fields that 
couple to gravity without any derivatives. We will call the latter Lagrange 
multiplier fields. Models of type II do not use Lagrange multiplier 
fields but consider only reference matter fields in their Lagrangians. 
Note that both types consider only scalar fields that are minimally coupled to
 gravity and hence the model of a conformally coupled scalar field in \cite{26a}
 is not considered in the classification here.
 \\
Common to both types of Lagrangians is that the non Lagrange 
multiplier fields enter the Lagrangian without a potential, that is,
it depends only on the first derivatives of the fields. Those fields are used to construct
the gauge invariant expressions of the remaining fields (geometry and standard matter) and as far as the final observables
are constructed those fields have been absorbed. This can be understood likewise to the case of the Higgs mechanism 
of the standard model where three out of the four Higgs fields are 'eaten' by the boson fields yielding to a
gauge invariant description of massive vector bosons. The main difference to the standard Higgs mechanism is that here
all reference field considered in the models are 'eaten' by the other degrees of freedom, 
whereas in the ordinary Higgs mechanism one scalar field remains
in the theory and is not absorbed.
The reason why no potentials of the reference fields are considered is  that one wants the physical Hamiltonian to be independent 
of the additional scalar fields introduced because those will play the role of the reference time and spatial reference points.
 Hence without a potential, on achieves that the physical Hamiltonian is then 
not explicitly dependent of the reference time and thus the reduced physical
system becomes conservative which has obvious advantages.
 In principle,
however, also potential terms could be treated if one gives up on the 
requirement of a conservative system. We will review both types of models I and II
below and point out some of their most important physical differences. 

\subsection{Type I}
\label{s2.1}

The first class of models are those 
considered in \cite{0,1,2} and we will briefly discuss their Lagrangians here. The famous Brown -- Kucha\v{r} timelike dust 
Lagrangian \cite{0} is given, in our notation, by
\be \label{2.1}
{\cal L}_{{\rm TD}}=-\frac{1}{2}\sqrt{|\det(g)|}\;\rho\;
[g^{\mu\nu} U_\mu U_\nu+1],\;\;U_\mu=\nabla_\mu T+W_j\;\nabla_\mu S^j
\ee
which depends on two pairs of respectively 4 scalar fields $(\rho\ge 0,W_j)$ 
and 
$(T,S^j)$ respectively. Here latin indices $j,k,..$ run from one 1 to 3 and greek indices $\mu,\nu,...$ run from 0 to 3. We choose our 
signature convention for the spacetime metric tensor $g_{\mu\nu}$ to be 
$(-,+,+,+)$. The Bi{\v c}ak -- Kucha{\v r} null dust Lagrangian \cite{1} is 
given by 
\be \label{2.2}
{\cal L}_{{\rm ND}}=-\frac{1}{2}\sqrt{|\det(g)|}\;
g^{\mu\nu} U_\mu U_\nu,\;\;U_\mu=W_j\;\nabla_\mu S^j
\ee
and results from (\ref{2.1}) by setting $\rho=1,\; T=0$ and dropping the 
cosmological 
constant term $-\rho\sqrt{|\det(g)|}/2$. The opposite limit was 
taken in a discussion in the appendix of \cite{0} where one sets $W_j=0$ and thus drops the dependence 
of (\ref{2.1}) on $S^j$ which results in 
\be \label{2.3}
{\cal L}_{{\rm NRD}}=-\frac{1}{2}\rho\sqrt{|\det(g)|}\;
[g^{\mu\nu} U_\mu U_\nu+1],\;\;U_\mu=\nabla_\mu T
\ee
This possibility of non rotational dust can be understood as a special case of the timelike dust model.
\\
The variables $(\rho,T)$ and $(W_j,S^j)$ appear symmetrically 
in (\ref{2.1}) in the sense that if we write $\lambda_0:=\sqrt{\rho},\;
\lambda_j:=\sqrt{\rho}W_j$ and $S^0:=T$ then (\ref{2.1}) becomes 
\be \label{2.4}
{\cal L}_{{\rm TD}}=-\frac{1}{2}\sqrt{|\det(g)|}\;
[g^{\mu\nu} U_\mu U_\nu+1],\;\;U_\mu=\lambda_\alpha \nabla_\mu S^\alpha
\ee
The Gaussian dust model suggested in \cite{2} treats them unsymmetrically 
and can be written in the form 
\be \label{2.5}
{\cal L}_{{\rm GD}}=-\frac{1}{2}\rho\sqrt{|\det(g)|}\;
[g^{\mu\nu} (\nabla_\mu T)\;(\nabla_\nu T)+1]
-\sqrt{|\det(g)|}\;g^{\mu\nu}(\nabla_\mu T) V_\nu,\;\;V_\mu=W_j \nabla_\mu S^j
\ee

\subsection{Type II}
\label{s2.2}

The first type of model of this class is simply the massless Klein-Gordon
action considered in \cite{4a} and whose canonical structure was discussed in detail in \cite{4}. It leads, for general 
reasons that we review below, to an Abelian algebra of Hamiltonian 
constraints. As pointed out in \cite{8}, the mechanism responsible for 
this is by far not restricted to the Klein Gordon action but works for 
a general Lagrangian of the form 
\be \label{2.6}
{\cal L}_{{\rm S}}=\sqrt{|\det(g)|}\;L(I),\;\;I=-\frac{1}{2} g^{\mu\nu}
[\nabla_\mu T]\;[\nabla_\nu T]
\ee
where $L$ is any function of the argument indicated.
The first observation is that (\ref{2.6}) in contrast to the 
type I Lagrangians depends on only a single scalar field $T$ rather than 
several ones $T,S^j$. The reason for this is that when one simply adds 
Lagrangians of the type (\ref{2.6}) or considers more generally 
Lagrangians depending on $L(I_1,..,I_N)$ where $I_k$ is as in 
(\ref{2.6}) but $T$ replaced by a field $T_k$, then the physical Hamiltonian 
no longer deparametrises. It is for this reason that the models of type I 
either do not depend on the $S^j$ or if they do then only in the combination
$V_\mu=W_j \nabla_\mu S^j$ involving the Lagrange multipliers. We will see
below why this is the case.

In any case, when we assume that observables are constructed
using the reference matter fields and no geometric degrees of
freedom we see that models of type II in contrast to some models of type I
cannot lead to a fully reduced phase space at the classical level
but only a reduction with respect to the Hamiltonian constraint
can be performed.

\subsection{Poisson commuting Hamiltonian constraints}
\label{s2.3}
One particular feature of the models in \cite{0,1,2,3,4a,4} is that 
they all involve Hamiltonian constraints that satisfy an Abelian algebra.
In the models in \cite{0,4a,4} one is furthermore able to rewrite the Hamiltonian constraints in deparametrised form
, that is in the form $P+h=0$ where $P$ is the momentum conjugate to the time reference 
field and $h$ is a function of all variables but the reference fields of density weight one, also 
called Kucha\v{r} density. As a consequence it follows immediately that also the Kucha\v{r}
densities $h$ mutually commute, that is $\{h(x),h(y)\}=0$ for all points $x,y$ in the spatial hypersurface. For the reason that in the model 
in \cite{4a,4} the spatially diffeomorphism constraint is not rewritten in a form linear in momenta
of reference field, the conclusion of strongly commuting Kucha\v{r} densities $h$ does not immediately follow
 but the algebra could close weakly, that is up to certain combinations of the spatial diffeomorphism constraint.
However, as proven in \cite{4} the corresponding Kuchar densities $h$ of this model also commute strongly.
\\
Those Kucha\v{r} densities 
are particular functions 
built from $Q=\sqrt{\det(q)},\;D=q^{ab} C_a C_b, \; C$ where 
$q_{ab}$ denotes the spatial three metric and $C_a,C$ respectively the 
non dust contributions to the spatial diffeomorphism and Hamiltonian 
constraints respectively (provided that matter couples only to the metric 
but not to its derivatives so that the hypersurface deformation algebra holds
\cite{26}). In general, rewriting the Hamiltonian 
constraints in this deparametrised form involves solving algebraic equations which leads to 
branches of the phase space labelled by the choice of certain signs, in this 
sense the description is only local on phase space and restricted to a single 
branch. 
\\
The existence of the Kucha\v{r} densities begs for the question whether one cannot
 obtain all functions $h$ without going through an action principle. This problem was
solved in \cite{6}. Following \cite{6} we introduce the density zero scalars $d=D/Q^2,\;c=C/Q$
and display a density $w$ Kucha{\v r} density in the form $h=Q^w K(c,d)$. 
Then the infinitely many equations $\{h(x),h(y)\}=0\;\forall\; x,y$ reduces, 
using the hypersurface 
deformation algebra, to the single first order PDE \cite{6}
\be \label{2.7}
\frac{w}{2} K\;\frac{\partial K}{\partial d}=d\;
\left(\frac{\partial K}{\partial d}\right)^2-
\frac{1}{4} \left(\frac{\partial K}{\partial c}\right)^2
\ee
A general integral can be obtained in terms of $S=\ln(K)$ by writing 
(\ref{2.7}) in the form 
\be \label{2.8}
-\frac{w}{2} \frac{\partial S}{\partial d}+d\;
\left(\frac{\partial S}{\partial d}\right)^2=
\frac{1}{4} \left(\frac{\partial S}{\partial c}\right)^2=:a^2
\ee
which allows to separate the variables $S(c,d)=S_1(c)+S_2(d)$
and to reduce (\ref{2.8}) to two simple quadratures which can be written
in closed form and involve two arbitrary parameters. The complete 
integral can then be obtained by the envelope construction. 

All solutions that were obtained by considering models of type I or II 
are of course described by (\ref{2.7}). One can ask the converse question
how to build a Lagrangian of type I or II respectively 
which reproduces a solution of (\ref{2.7}). This was analysed in \cite{7} and 
\cite{8} respectively. In \cite{7} it is shown that for all solutions of 
(\ref{2.7}) a type I action involving a single
scalar field $T$ with non derivative coupling of the metric 
and a single Lagrange multiplier field $\rho$ can be found reproducing it.
Remarkably, all solutions of (\ref{2.7}) satisfying certain reality 
conditions (such as the positivity of the kinetic term) can be obtained from 
an action of the form 
\be \label{2.9}
{\cal L}'_{{\rm NRD}}=-\frac{1}{2}\sqrt{|\det(g)|}\;
[\rho \;g^{\mu\nu} U_\mu U_\nu+\Lambda(\rho)],\;\;U_\mu=\nabla_\mu T
\ee
That is, it is of the non rotating dust type just that the cosmological 
constant term is allowed to be a general function of the Lagrange multiplier.
It has to satisfy an ODE that is matched to the two ODE's in (\ref{2.8}).
By contrast, the class of solutions of (\ref{2.7}) that one obtains from the 
type II models is more restricted. One finds from the Legendre transform  ($P=\partial L/\partial \dot{T}$)  the following expression for $p^2:=(P/Q)^2$ as shown in \cite{7}
\be \label{2.10}
p^2:=\left(\frac{P}{Q}\right)^2=(L'(I))^2(I+q^{ab} T_{,a} T_{,b})
=(L'(I))^2\left(I+\frac{d}{p^2}\right)
\ee
In the second step we used the spatial diffeomorphism constraint and applied the Brown-Kucha{\v r} mechanism, that is using the relation
$q^{ab} T_{,a} T_{,b}=D/P^2$ in order to replace the second term in the brackets above.
The Hamiltonian constraint is given by 
\be \label{2.11}
C^{{\rm tot}}=Q\left(c+\frac{p^2}{L'(I)}-L(I)\right)
\ee
where $I$ solves (\ref{2.10}). To check whether a given solution $p=-K(c,d)$ 
of (\ref{2.7}) corresponds to a Lagrangian $L(I)$ we insert $p=-K(c,d)$ into 
(\ref{2.10}) and solve the resulting equation for $c=c(I,d)$. Then 
(\ref{2.11}) must become the identity
\be \label{2.12}
0=c(I,d)+\frac{K(c(I,d),d)^2}{L'(I)}-L(I)
\ee
which is an ODE for $L$ provided the explicit 
dependence of (\ref{2.12}) on $d$ drops out.

\subsection{Global Parametrisation of Theory Space}
\label{s2.4}

We now combine these known results and write down an action including up 
to 8 scalar fields and several parameters that describes all the models so far 
considered. Since 
the models of type II are contained in the set of models of type I in 
the fashion described we 
focus on the former set. Consider 
\be \label{2.13}
{\cal L}_D=-\frac{1}{2}\sqrt{|\det(g)|}\left(g^{\mu\nu}\left[ 
\rho\;(\nabla_\mu T)\;(\nabla_\nu T)+A(\rho) V_\mu V_\nu+2 B(\rho)
(\nabla_\mu T) V_\nu\right]+\Lambda(\rho)
\right),\;\;V_\mu=W_j \nabla_\mu S^j
\ee
where $A,B,\Lambda$ are arbitrary functions of the field $\rho$ (we could
also have considered an arbitrary function $F(\rho)$ rather than $\rho$ as 
the coefficient of the $(\nabla T)^2$ term but this can be absorbed by a 
field redefinition $F(\rho)=\rho'$. Below 
we list the choices of these functions corresponding to the models
reviewed in section (\ref{s2.1})
\be \label{2.14}
\begin{array}{cc}
{\cal L}_{TD} & A=B=\Lambda=\rho\\
{\cal L}_{ND} & A=1,\;\rho=B=\Lambda=0\\
{\cal L}_{NRD} & A=B=0,\;\Lambda=\rho\\
{\cal L}_{GD} & A=0,\;B=1,\Lambda=\rho
\end{array}
\ee
Among these models we find many new ones such as $A=1,\;B=0,\;\Lambda=\rho$
and of course all the models that one obtains by choosing $A,B,\Lambda$ of a 
more general than linear form (notice however that unless $A=0$ we may 
absorb $A$ into the $W_j$ and have only $B$ as a free function at our 
disposal. We keep both $A,B$ so that we can treat the general case in a 
unified form).

\subsection{General Hamiltonian Analysis of Theory Space}
\label{s2.5}

The general Hamiltonian analysis of the action corresponding to the 
sum of (\ref{2.13}) and the standard matter and geometry contributions
proceeds roughly as follows (we consider the case that $\rho\not=0$ and 
that not both of $A,B$ vanish, more singular cases can be treated similarly, 
see \cite{15} for all the details in the case $A,B\not=0$):\\
Computing the momenta $P,P_j$ conjugate to $T,S^j$, the momenta $\pi,\pi^j$
to $\rho, W_j$ and the momenta $\Pi,\Pi_a$ to lapse and shift functions 
$N,N^a$ in the 3+1 split of the action, we discover the primary constraints 
$z=\pi=0,\;z^j=\pi^j=0,\;Z=\Pi=0,\;Z_a=0$ as well as another set of 3
(2) linearly independent constraints $\zeta^j=0$ for $A\not=0$ ($A=0$)
which involves only $P, P_j, W_j$ ($P_j, W_j$), in both cases demand 
that $P_j W_k-W_k P_j=0$ and which come from the fact that one cannot 
solve for 3 (2) of the velocities $\dot{S}^j$. Altogether the canonical 
Hamiltonian depends on 11 (10) undetermined 
velocities. The stability analysis of the 
primary constraints with respect to the canonical Hamiltonian 
$H_{{\rm can}}$ yields that one can solve  
\be \label{2.15}
\{\pi^j,H_{{\rm can}}\}=0,\;\;j=1,..,3\, (2)\quad\quad
\{\zeta^j,H_{{\rm can}}\}=0,\;j=1,..,3\, (2)
\ee
for 6 (4) of the velocities and that there are the 5 (6) 
secondary constraints 
\be \label{2.16}
s=\frac{\partial H_{{\rm can}}}{\partial \rho}=0,\,\,
C^{{\rm tot}}=\frac{\partial H_{{\rm can}}}{\partial N}=0,\,\, 
C^{{\rm tot}}_a=\frac{\partial H_{{\rm can}}}{\partial N^a}=0,\quad
(K=\{\pi^3,\partial H_{{\rm can}}\}=0)
\ee
The secondary constraints can be stabilised by solving for 1 (2) of the 
remaining unfixed velocities so that altogether 7 of 11 (6 of 10) have been 
fixed, leaving in both cased $\dot{N},\;\dot{N}^a$ undetermined. Altogether
5 (6) secondary constraints were found leading to altogether 16 constraints.

One finds that the 4 pairs $(z,s),(z^j,\zeta^j);\;j=1,..,3$ 
($(z,s),(z^j,\zeta^j),\;j=1,2; (z^3,K)$) form a second class 
system while
linear combinations of $Z, Z_a, C^{{\rm tot}}, C^{{\rm tot}}_a$ with 
the second class constraints forms a first class set.
The Dirac bracket
between functions independent of $\rho,W_j,N,N^a$ and their conjugate
momenta reduces to the original Poisson bracket. This follows 
from the fact that such functions have vanishing Poisson brackets 
with the constraints $z,z^j$ which mutually commute among themselves.
The inverse of the Dirac matrix between the second class constraints 
is therefore such that the difference between the difference between
the Poisson bracket and the Dirac bracket involves at least one 
Poisson bracket with one of the $z,z^j$. See section \ref{s4} or \cite{15}
for more details. One then solves the second 
class constraints strongly thereby eliminating $\rho,W_j$ from the canonical 
Hamiltonian altogether. The result is the same as if one had eliminated 
right from the beginning $\rho,W_j$ by using 
$\partial H_{{\rm can}}/\partial \rho=
\partial H_{{\rm can}}/\partial W_j=0$ and dropping the primary constraints
from the Hamiltonian. This follows because these four equations are just
four of the second class constraints. 

The canonical Hamiltonian is then a linear combination of the 
first class constraints $C^{{\rm tot}},\;C^{{\rm tot}}_a$ which besides 
geometry and standard matter degrees of freedom, which we denote 
collectively by $(q,p)$, now only 
depend on $(T,P),\;(P_j, S^j)$ if not both $A,B$ vanish and otherwise only
on $(T,P)$. Crucially, since in the first case the Hamiltonian depended 
only on the combination $W_j S^j_{,a}$ and since $W_j\propto P_j$ on the 
constraint surface of the second class constraints, the canonical Hamiltonian 
eventually depends only on the combination 
$P_j S^j_{,a}$ as far as the $S^j$ dependent
terms are concerned. This combination, however, is weakly equal to 
$-(C_a+P T_{,a})$ (see below) where $C_a$ is the contribution of geometry and 
standard matter to the spatial diffeomorphism constraint. It is for that 
reason (extended 
Brown -- Kucha{\v r} mechanism) that this particular combination was chosen 
in the original Lagrangian. In both cases, the explicit dependence on $T$ 
in the Hamiltonian constraint is through $T_{,a}$ only and the mechanism 
just displayed makes sure that this remains true also when substituting
$P_j S^j_{,a}$. 

The remaining analysis of the system then proceeds as follows. 
It follows from the above that in the first case the constraints 
can be written in the equivalent form 
\be \label{2.17}
C^{{\rm tot}}=P+h(T,q,p),\;\;
C_a^{{\rm tot}}=P T_{,a}+P_j S^j_{,a}+C_a(q,p)
\ee
while in the second one has 
\be \label{2.18}
C^{{\rm tot}}=P+\tilde{h}(T,q,p),\;\;
C_a^{{\rm tot}}=P T_{,a}+C_a(q,p)
\ee
In the timelike dust model the dependence of $h$ on $T$ even drops out 
completely while in (\ref{2.18}) the fact that $\tilde{h}$ depends only on 
$q^{ab} T_{,a} T_{,b}\approx q^{ab} C_a C_b/P^2$ can be used to further 
massage the constraints into the form 
\be \label{2.19}
C^{{\rm tot}}=P+h(q,p),\;\;
C_a^{{\rm tot}}=P T_{,a}+C_a(q,p)
\ee
In all other models, $h$ keeps an explicit dependence on $T_{,a}$. 
Nevertheless, in all models the Hamiltonian constraints in 
(\ref{2.17}), (\ref{2.18}) and 
(\ref{2.19}) strongly Poisson commute. This is because they are 
first class by construction but since the Poisson bracket eliminates the 
dependence on $P$, the bracket must actually vanish identically in the four reference
 field case ($\rho\not=0$ and not both $A,B$ vanish).
Notice, however, that this only implies that also the $h$ strongly 
Poisson commute if they are independent of $T$. In the single reference field it is more complicated 
to show that the $h$ strongly commute as has been discussed in \cite{8} and \cite{4}.
\\
\\
In the case of model (\ref{2.17}) one can now perform a symplectic reduction
of the spatial diffeomorphism constraint by pulling back all tensors 
and spinors by the diffeomorphism $x\mapsto \sigma^j=S^j(x)$ 
which is a canonical transformation \cite{0,15} to the effect that 
all degrees of freedom except $P_j,S^j$ remain canonical pairs when 
expressed in the new frame while in the new frame the momentum conjugate to 
$S^j$ becomes $([\partial S/\partial x]^{-1})^a_j C^{{\rm tot}}_a$ so 
that this canonical pair drops out from our attention. The remaining
degrees of freedom are spatially diffeomorphism invariant when expressed in 
this frame. We will denote them by $(q',p',T',P')$ in order to distinguish them 
from $(q,p,T,P)$. In case of 
(\ref{2.19}), spatial diffeomorphism invariant quantities must be constructed
by other means. We will also pull back $C^{{\rm tot}}(x)$ which then 
becomes the same function $C^{{\rm tot}\prime}(\sigma)$ 
of $q',p'T',P'$ at $\sigma$ 
as $C^{{\rm tot}}(x)$ was of $q,p,T,P$ at $x$.

The final reduction of the system now employs the projector formula 
discovered in \cite{9,10}. See for a review in \cite{15} in the notation 
used here.
Let $F',F$ be spatially diffeomorphism 
invariant functions on the phase space respectively 
which depend only on $(q',p')$ in case 
of (\ref{2.17}) and only on $(q,p)$ in case of (\ref{2.19}) respectively.
Let 
\be \label{2.20}
O'_{F'}(\tau):=(\exp(\{C^{{\rm tot}\prime}[g'],.\})\cdot F')_{g'=T'-\tau};\; 
O_F(\tau):=(\exp(\{C^{{\rm tot}}[g],.\})\cdot F)_{g=T-\tau};\; 
\ee
The notation means that we compute the Hamiltonian flow of the constraint
indicated when smeared with a numerical test function $g'$ or $g$ and 
then set $g'=T'-\tau$ or $g=T-\tau$. We cannot directly insert $T-\tau$
into the exponent due to the phase space $T$ dependence. 
Formula (\ref{2.20}) is rather remarkable in several aspects. First of 
all (\ref{2.20}) strongly Poisson commutes with all constraints. Secondly,
we have the equal time $\tau$ Poisson brackets
\be \label{2.21}
\{O'_{F'}(\tau),O'_{G'}(\tau)\}=O'_{\{F',G'\}^\ast}(\tau)
\ee 
where $\{.,.\}^\ast$ is the Dirac bracket with respect to the second class
pair $T,C^{{\rm tot}\prime}$ (similar statements hold for the unprimed 
quantities). For quantities $F',G'$ independent of $P'$ it reduces to 
the normal Poisson bracket. It follows that if $F',G'$ are conjugate so 
are $O'_{F'}(\tau),\;O'_{G'}(\tau)$. Secondly we have due to the 
mutual strong commutativity of the $C^{{\rm tot}\prime}(\sigma)$
\ba \label{2.22}
\frac{d}{d\tau} O'_{F'}(\tau) 
&=&
\sum_{n=1}^\infty \frac{1}{(n-1)!}\int\; d^3\sigma_1 .. \int\; d^3\sigma_n
[\tau-T'(\sigma_1)]..[\tau-T'(\sigma_{n-1})]\;
\{C^{{\rm tot}}(\sigma_1),\{..\{C^{{\rm tot}}(\sigma_n),F'\}..\}\}
\nonumber\\
&=& O'_{\{C^{{\rm tot}\prime}[1],F'\}}(\tau)
= O'_{h'[1],F'\}}(\tau)
= O'_{\{h'[1],F'\}^\ast}(\tau)
\nonumber\\
&=& \{O'_{\{h'[1]}(\tau),O'_{F'}(\tau)\}
\ea
where in the second step we realised the most inner Poisson bracket as the 
one between 
$C^{{\rm tot}\prime}[1]=\int\;d^3\sigma C^{{\rm tot}\prime}(\sigma)$ and $F'$,
in the third we used independence of $F'$ on $T'$ and that 
$C^{{\rm tot}\prime}=P'+h'$, in the fourth we used independence of both 
$h',F'$ of $P'$ and in the last we used (\ref{2.21}).
The next property we need is that $F'\mapsto O'_{F'}(\tau)$ 
preserves the multiplicative and additive structure on the Abelian 
algebra of phase space 
functions, that is, 
\be \label{2.23}
O'_{F'}(\tau)=F'(O'_{q'}(\tau),O'_{p'}(\tau),O'_{T'}(\tau))
\ee
if $F'=F'(q',p',T')$ depends on $q',p',T'$ only. 
Now, the observables associated to the time reference field is given by 
$O'_{T'}(\tau)=\tau$ where $\tau$ is a spatial constant. Likewise we obtain for the spatial derivative of $T$ with 
respect to the (dust) spatial coordinates $T_{,\sigma^j}$
 the following observable  $O'_{T'{,\sigma_j}}(\tau)=\partial_{\sigma^j}\tau=0$. Accordingly,
since $h'$ only depends on the derivatives of $T'$ because we assumed that we do not consider
potential terms of the reference matter fields in all models and not on $T'$ itself
we can define the intermediate Hamiltonian by 
\be \label{2.24}
H':=\int\; d^3\sigma\; h'(T',q',p')_{T'=0}
\ee
Then (\ref{2.22}) becomes
\be \label{2.25}
\frac{d}{d\tau} O'_{F'}(\tau)=\{O'_{H'}(\tau),O'_{F'}(\tau)\}
=O'_{\{H',F'\}^\ast}(\tau)=O'_{\{H',F'\}}(\tau)
\ee
where independence of $H'$ of $P'$ was used again.
The properties of $F'$ that were used in this derivation is that $F'$ 
is independent of both $T',P'$. Formula (\ref{2.25}) therefore in particular
applies to $F'=H'$ whence 
\be \label{2.26}
\frac{d}{d\tau} O'_{H'}(\tau)=0
\ee
is in fact $\tau$ independent.
It follows that (\ref{2.25}) can be rewritten 
as
\be \label{2.27}
\frac{d}{d\tau} O'_{F'}(\tau)=\{\hat{H},O'_{F'}(\tau)\}
\ee
where we have identified the 
physical, fully gauge invariant Hamiltonian
\be \label{2.28}
\hat{H}:=O'_{H'}(0)=O'_{h'[1]}(0)
\ee
and in the second step we have made again use of $O'_{T'}(0)=0$. 
\\
\\
Alternatively, one can derive the fact that the physical Hamiltonian $\hat{H}$ 
is not depending on time in the models considered as follows:
\\
In case the function $F'$ depends explicitly on $\tau$ the formula in (2.22) needs to be 
modified and we obtain
\be
\label{Fexpl}
\frac{d}{d\tau} O'_{F'}(\tau) 
= \{O'_{h'[1]}(\tau),O'_{F'}(\tau)\} +\frac{\partial O'_{F'}}{\partial\tau}
\ee
Now, let us consider the case that $h'[1]$ depends on $q',p',T'$ but not on $P'$, then we have
\be
h'[1]=\int d^3\sigma h'(T',q',p')
\ee
Using that the application of the projector $F'\mapsto O'_{F'}(\tau)$ preserves the
multiplicative and additive structure on the Abelian algebra of phase space
functions, we obtain
\be
\hat{H}=O'_{h[1]}(\tau)=\int d^3\sigma h'(O'_{T'},O_{q'},O_{p'})
\ee
Assuming that we do not consider potential terms of the reference fields, as it is done
in all models considered in this paper, they occur only
with spatial derivatives. Now the observable associated to the time reference field
is just given by $O'_{T'}(\tau)=\tau$. Since $\tau$ does not depend on the spatial (dust)
coordinates we further get $O'_{T'{,\sigma^j}}(\tau)=\tau_{,\sigma^j}=0$.
Consequently, in this case the physical Hamiltonian is of the form
\be
O'_{h[1]}(\tau)=\int d^3\sigma h'(O_{q'}(\tau),O_{p'}(\tau))
\ee
and has no explicit $\tau$-dependence. Knowing this, we can immediately see that it
is constant with respect to the $\tau$-evolution using (\ref{Fexpl})
\be
\frac{d}{d\tau} O'_{h'[1]}(\tau) 
= \{O'_{h'[1]}(\tau),O'_{h'[1]}(\tau)\} +\frac{\partial O'_{h'[1]}}{\partial\tau}=0
\ee
Therefore we can use any value of $\tau$ for the fully gauge invariant Hamiltonian $\hat{H}$
and define
\be
\label{Hphys}
\hat{H}:=O'_{h'[1]}(0)
\ee
\\
\\
Using (\ref{2.23}) these are precisely the Hamiltonian equations of motion
with respect to the Hamiltonian (\ref{Hphys}). Notice that unless 
$h'_{|T'=0}$ is a Kucha{\v r} density, $H'$ itself is not invariant 
under the Hamiltonian flow of generated by $C^{{\rm tot}\prime}$, 
whence in this case the projector on $H'$ in (\ref{2.24}) is not 
trivial and $H'$ must be replaced by (\ref{2.28}). Also, 
unless $h'_{|T'=0}$ is a Kucha{\v r} density, it will not be preserved by 
$\hat{H}$ whence $\hat{H}$ in general has only the three sets of local 
Noether charges 
$C'_j$ which arises as the pull back of $C_a$ by the diffeomorphism $S^j$
(notice that $h',C^{{\rm tot}\prime}$ are a scalar densities of 
weight one in $\sigma$ space and that $\tau -T$ is a scalar so that 
(\ref{2.20}) maps the diffeomorphism invariant quantity $h'[1]$ to a 
diffeomorphism invariant quantity.).

We now reformulate in manifestly gauge invariant form. Consider 
the Dirac observables $\hat{q}=O'_{q'}(0), \; \hat{p}=O'_{p'}(0),\;\hat{H}=O'_{h'[1]}(0)=h'[1](\hat{q},\hat{p})=H'(\hat{q},\hat{p})$, where we assume that $h[1]$ includes only derivatives of the time reference field.
Then (\ref{2.21}) assures that 
Poisson brackets 
between functions of the hatted quantities can be computed by computing 
the brackets between the same functions of the primed quantities and then 
evaluated at the hatted quantities.

\section{Non Rotational Dust}
\label{s3}

We carry out some of the details sketched in section (\ref{s2.5}) for 
the choice $A=B=0,\Lambda=\rho$ from a manifestly gauge invariant
viewpoint and compare the results with \cite{22}. 
Of course, almost everything we say here is already contained in 
\cite{0,19}. We will show that this model is contained as a subsector 
in the model \cite{19}, both classically and in the quantum theory and also
 explain why proposal for the quantum theory in \cite{22} cannot describe
 the physical sector of the model.\\
\\
In the case of non rotational dust the primary constraints are $\pi=\Pi=\Pi_a=0$ and the canonical Hamiltonian 
is given by 
\be \label{3.1}
H_{{\rm can}}=N^a C_a^{{\rm tot}}+N\left(C
+\frac{1}{2}\left(\frac{P^2}{\rho\sqrt{\det(q)}}+
\rho\sqrt{\det(q)} [q^{ab} T_{,a} T_{,b}+1]\right)\right)+v\pi+U\Pi+U^a\Pi_a
\ee
with undetermined velocities $v,U,U^a$. 
Stability of the primary constrains requires
\ba \label{3.2}
C_a^{{\rm tot}} &=& P T_{,a}+C_a=0
\nonumber\\
C^{{\rm tot}} &=& C+\frac{1}{2}\left(\frac{P^2}{\rho\sqrt{\det(q)}}+
\rho\sqrt{\det(q)} [q^{ab} T_{,a} T_{,b}+1]\right)=0
\nonumber\\
s &=& -\frac{P^2}{\rho^2\sqrt{\det(q)}}+
\sqrt{\det(q)} [q^{ab} T_{,a} T_{,b}+1]=0
\ea
Since $\{\pi,s\}\not=0$, the constraints $S,\pi$ form a second class pair
and stability of $s$ can be 
ensured by choosing $v$. We add to $C^{\rm tot}_a$ the term $\pi \rho_{,a}$
proportional to the second class constraint $\pi$. Then this enlarged 
constraint generates spatial diffeomorphisms on all variables contained
in $C^{{\rm tot}}_a,\; C^{{\rm tot}}$ and thus preserves these secondary 
constraints. The smeared $C^{{\rm tot}}$ Poisson commute to a smeared 
spatial diffeomorphism constraint 
according to the hypersurface deformation algebra
because, up to the factors of $\rho$, the dust contribution $C^D$ 
to the Hamiltonian constraints coincides with that of a massless Klein-Gordon 
field plus a cosmological constant term, $\{C,C^D\}$ 
does not involve Poisson brackets between derivatives    
of fields and $C^D,C^D$ lets its derivatives act on the smearing fields only
so that the factors of $\rho,1/\rho$ in $C^D$ cancel for the same reason as 
$\sqrt{\det(q)},1/\sqrt{\det(q)}$. It follows that there is only one 
second class pair. The corresponding Dirac bracket $\{f,g\}^\ast$ 
differs from the 
Poisson bracket by terms that involves terms of the form 
$\{\pi,f\}\{\pi,g\},\;
\{s,f\}\{\pi,g\},\;
\{\pi,f\}\{s,g\}$ since $\{\pi,\pi\}=0\not=\{s,s\},\{s,\pi\}$ (the inverse
of the Dirac matrix $\Delta_{ij}=\{s_i,s_j\},\;s_1=s,\;s_2=\pi$
has the zeroes in complementary places as compared to the Dirac matrix
itself). Hence, the Dirac bracket of 
phase space functions independent of $\pi,\rho$ coincides with the 
Poisson bracket. We solve the second class constraints strongly and obtain
\be \label{3.3}
\pi=0,\;\rho^2=\frac{P^2}{\det(q)[1+q^{ab} T_{,a} T_{,b}]}
\ee
Choosing a sign $\epsilon$ for $\rho$ we find that the canonical Hamiltonian 
is a linear combination of first class constraints
\ba \label{3.4}
C_a^{{\rm tot}} &=& P T_{,a}+C_a=0
\nonumber\\
C^{{\rm tot}} &=& C+\epsilon |P|\sqrt{1+q^{ab} T_{,a} T_{,b}}
\ea
This can be massaged into 
\be \label{3.5}
C^{{\rm tot}} = C+\epsilon \sqrt{P^2+q^{ab} C_a C_b}
\ee
using the first equation in (\ref{3.4}). On the constraint surface, $C$ 
has the sign $-\epsilon$ (unless all quantities vanish). 
Inverting for 
$|P|$ we find 
\be \label{3.6}
|P|^2=C^2-q^{ab} C_a C_b
\ee
which is constrained to be non negative. Thus, choosing a sign $\delta$ 
for $P$ we may replace (\ref{3.5}) by
\be \label{3.7} 
C^{{\rm tot}}=P-\delta\sqrt{C^2-q^{ab} C_a C_b}=P-\delta h
\ee
In the limit $q^{ab} C_a C_b\to 0$ (\ref{3.5}) and (\ref{3.7}) become 
$C+\epsilon|P|=P-\delta |C|=0$ which is consistent for either 
choice of sign $\epsilon,\delta$. It is customary to choose 
$\epsilon=1=-\delta$ (this is one of the four possible subsectors of 
phase space that one must pick in these local considerations). 
As $H$ no longer depends on $T$ it must be 
a Kucha{\v r} density which maybe checked explicitly. We stress 
here that that in this model, as one can see explicitly from  
(\ref{3.4}) or (\ref{3.5}), 
the sign of $C$ is constrained, specifically $-\epsilon$ is the sign of $C$.
In particular, for $\epsilon=1$ (positive kinetic energy term), 
$C$ is constrained to be negative. 

The physical Hamiltonian becomes by the general considerations of the 
previous section the manifestly positive expression (whatever choice of 
$\epsilon,\delta$ would have been made, (\ref{3.8}) has positive sign)
\be \label{3.8}
\hat{H}=\int \;d^3x \; h(x)
\ee
generating time evolution of functions of the form 
\be \label{3.9}
O_F(\tau)=[\exp(\{C^{{\rm tot}}[g],.\})\cdot F]_{g=T-\tau}
\ee
where $F$ is spatially diffeomorphism invariant and independent of $T,P$. 
The functions (\ref{3.8}) and (\ref{3.9}) are full Dirac observables 
of the theory and their algebra may be computed on the constraint surface
$C^{{\rm tot}}=C^{{\rm tot}}=0$. Since they are gauge invariant, their 
algebra may be evaluated on the gauge cut $T=\tau_0$ for some fixed 
$\tau_0=const.$, say $\tau_0=0$ within that constraint
surface. On that gauge cut we have $C^{{\rm tot}}_a=C_a=0$ and $h=|C|$
as well as $O_F(\tau)=\exp([\tau-\tau_0]\{h[1],.\})\cdot F$, in particular
$O_F(\tau_0)=F$. The generator 
of time evolution is evidently still given by $H=h[1]$ in (\ref{3.8}) but 
on the gauge cut its Hamiltonian vector field reduced to that of $|C|[1]$
as follows from 
\ba \label{3.10}
\{H,f\}
&=& \int\; d^3x \; \frac{1}{2h}(2C\{C,f\}-2q^{ab} C_b\{C_a,f\}
-C_a C_b\{q^{ab},f\})_{C_a=0}
\nonumber\\
&=& \int\; d^3x \; {\rm sgn}(C)\;\{C,f\}
=\{|C|[1],f\}
\ea
A different way to see this is to start from the formulation (\ref{3.4}) 
and to write the Hamiltonian constraint in the equivalent form
\be \label{3.11}
C^{{\rm tot}}=P+\frac{|C|}{\sqrt{1+q^{ab} T_{,a} T_{,b}}}=:P+\tilde{h}
\ee
where the  sign choice $\epsilon=1=-\delta$ has been adopted.  
The constraints (\ref{3.11}) are still Abelian and the machinery sketched 
in the previous section applies but $\tilde{h}$ is no Kucha{\v r} density 
any longer and $\tilde{h}[1]$ is no longer a Dirac observable. 
However, the theory sketched there shows that the physical 
Hamiltonian is given by (notice that $\tilde{O}_\cdot(\tau)$ is now defined 
in terms of $\tilde{h}$ while $O_\cdot(\tau)$ is defined in terms of $h$)
\be \label{3.12}
\hat{H}=\tilde{O}_{\tilde{h}[1]}(0)=\tilde{O}_{|C|[1]}(0)
\ee
since $O_{T_{,a}}(0)=0$. As the algebra of the $F$ and $|C|$ is isomorphic 
to the algebra of the $O_{F}(0)$ and $O_{|C|}(0)$ we arrive at the same 
picture at this fully gauge invariant level as in its gauge fixed version.
Yet a third way to see this is to start from (\ref{3.8}) and to make
use of the fact that $h$ is a Kucha{\v r} density. Then trivially
$H=O_{H}(0)=\hat{H}$. On the other hand in $O_{H}(0)$ we may replace 
$C_a$ appearing in $h=\sqrt{C^2-q^{ab} C_a C_b}$ by $C^{{\rm tot}}_a$ 
as $O_{T_,a}(0)=0$. Writing $\hat{C}=O_C(0),\;\hat{C}^{{\rm tot}}_a=
O_{C^{{\rm tot}}_a}(0)$ etc. we have 
\be \label{3.13}
\hat{H}=\int\; d^3x\; \sqrt{[\hat{C}]^2-\hat{q}^{ab} 
\hat{C}^{{\rm tot}}_a 
\hat{C}^{{\rm tot}}_b}
\ee
and now a calculation similar to (\ref{3.10}) reveals that on the constraint 
surface we may replace $\hat{H}$ by $O_{|C|[1]}(0)$. We conclude that 
on the constraint surface all three approaches are equivalent.\\
\\
Notice that without the 
additional fields $S^j$ spatially diffeomorphism functions $F$ are difficult 
to construct: Typically 
one will form them from scalar densities of weight one, built from the 
the fields $q,p$ and their spatial covariant derivatives. These are 
no longer elementary functions such as $q,p$ in terms of which a quantisation
is typically easy. 
The equal time 
Poisson algebra of the observables $O_F(\tau)$ is isomorphic to that 
of the $F$ since they do not depend on the $P,T$. 
However, since it is unknown how to construct a complete and independent 
set of such 
$F$ and thus how to write their algebra in closed form, it is not possible 
to use the $O_F(0)$ or equivalently the $F$ as a platform for quantisation
and thus a full reduced phase space quantisation is impossible for 
practical reasons. One thus has to resort to a hybrid scheme such as in
\cite{20} where one starts from the phase space unreduced with respect 
to the spatial diffeomorphism constraint and 
quantises the $O_f(0)$ whose algebra is 
isomorphic\footnote{In \cite{20} actually the $f$ were quantised and the 
projector formula $O_\cdot(\tau)$ was implemented in the quantum theory.
The results are equivalent.}
to that of the $f$. The spatial diffeomorphism constraint is then solved 
in the quantum theory following \cite{27}. However, while this 
equips the theory with a physical Hamiltonian $H$ (for the 
matter model considered in \cite{4a,4} rather than that model of non rotational dust considered as a special case of \cite{0} used in 
this section), the task 
of implementing the $O_F(0)$ remains to be performed and the reservations 
of \cite{21} are sustained.
\\
\\
Coming back to the model of this section that was studied in \cite{22}
 the spatial diffeomorphism constraint needs to be solved at the quantum level
 for the reason that likewise to \cite{20} only one (time) reference field is involved in that
 model. Hence in order to describe the physical sector of the quantum
 theory, the  same procedure as in \cite{20} must be adopted as far as the quantum 
theory is concerned and the physical Hamiltonian needs to be quantised on the
diffeomorphism invariant Hilbert space. However, the authors of \cite{22} suggest a different
procedure. Rather than quantising the theory and the physical Hamiltonian
$|C|$ on the diffeomorphism invariant Hilbert ${\cal H}_{{\rm diff}}$
space constructed in \cite{27}
and employed in \cite{20}, they suggest to use the algebraic quantum gravity
framework of \cite{19} and to quantise $q,p,|C|$ directly on the corresponding 
Hilbert space ${\cal H}_{{\rm AQG}}$. Since that Hilbert space is based on 
abstract rather than embedded graphs, their viewpoint seems to be that 
therefore the spatial diffeomorphism constraint (and by the above gauge
fixing viewpoint also the Hamiltonian constraint) can be considered 
as solved so that ${\cal H}_{{\rm AQG}}$ is actually the physical 
Hilbert space with the simplest possible physical Hamiltonian $C$ on it.
We make five comments about this:
\begin{itemize}
\item[1.] First of all, since really the physical Hamiltonian is $|C|$ rather
than $C$ there is secretly still a square root $|C|:=\sqrt{C^2}$ involved 
which makes $|C|$ no better than $h=\sqrt{C^2-q^{ab} C_a C_b}$ when quantising
the theory. However, note that there is no principal problem with the square 
root
because one can use the methods of \cite{23} to handle it.
The authors of \cite{22} could of course take the point of view that they 
fix the sign of $C$
classically and then quantise the corresponding part of the phase space in 
order to avoid 
the sign function at the quantum level. However, this is almost impossible 
to control in practical terms, since it requires detailed knowledge about 
the spectrum of $C$. The only practical way of ensuring the sign is 
to introduce the absolute value (that is a square root).
Alternatively, the authors of \cite{22} could argue to relax the 
typical energy conditions of classical General Relativity  
which however is not unproblematic. In any case, this discussion 
shows, that the claim that the physical Hamiltonian simplifies 
in the maximal sense and just reduces to $C$ deserves further discussion
By contrast, as we will discuss in the next section, there exists
  already a model \cite{2}, the Gaussian dust model, in which one can obtain 
physical Hamiltonian
   densities that simplify in the maximal sense and are just given by 
$C$ without the need to fix signs at 
   the classical level.

\item[2.] As noticed previously in the literature (see e.g. \cite{24} and 
references therein) the Hilbert space ${\cal H}_{{\rm AQG}}$ cannot be 
identified with the physical Hilbert space for  models where the 
spatial diffeomorphism constraint is not solved already classically. 
We repeat the argument here for completeness: \\
The physical interpretation of spin network functions in 
${\cal H}_{{\rm AQG}}$ whose abstract 
graphs are finite subgraphs of an infinite abstract graph is obtained 
by embedding them into the given spatial manifold $\sigma$ 
under consideration. 
Any such embedding $Y$ establishes an isomorphism, at least when $\sigma$ 
is compact, with a subspace of 
the kinematical Hilbert space ${\cal H}_{{\rm LQG}}$ of LQG. In that 
Hilbert space, spin network functions based on diffeomorphic graphs
are gauge equivalent. Hence must be their preimages in 
${\cal H}_{{\rm AQG}}$ under $Y$. For instance, if the infinite algebraic 
graph is cubic then all Wilson loop functions around plaquette loops 
should be identified. 

The physical sector of the quantum theory in AQG
 is determined by the kernel of the extended master constraint, which includes,
  for the reasons mentioned above, also a master constraint version 
  of the spatial diffeomorphism constraint at the algebraic level.
The quantisation of the (not necessarily algebraic graph-preserving) 
extended master constraint 
is presented in \cite{23} using techniques from \cite{32}. (Note that the continuum version of the spatial 
diffeomorphisms constraint used at the embedded level
has no counterpart at the algebraic level). Therefore, when solving the Hamiltonian 
constraint classically, as it is done in \cite{22}, one still needs to find the
solutions of the operator corresponding to the classical expression 
\be
\label{MAQG}
M:=\int d^3\sigma \frac{q^{ab}C_aC_b}{\sqrt{\det(q)}}
\ee
in order to solve the spatial diffeomorphism constraint at the quantum level and thus describe
the physical sector of the theory. The quantisation of such an operator
was presented in \cite{32,23}, however in the case of the model in \cite{22} this quantisation cannot be copied.
The reason is the following: Assume that one is able to find the solution space of the (corresponding) operator in (\ref{MAQG}).
Then this solutions space is not left invariant under the action of the physical Hamiltonian because $q^{ab}C_aC_b$ will not
commute with the physical Hamiltonian densities given by $C$ (or $|C|$ respectively when the sign is not fixed at the 
classical level). Therefore one needs to find a quantisation of 
$M$ 
such that the corresponding operator commutes with $C$ modulo 
$M$. Such a quantisation\footnote{Note that it is not enough 
to commute modulo $C_a$ which classically is trivially the case because 
$C_a$ in contrast to $M$ is not a well defined operator.} 
so far does not exist
in the literature and also has not been worked out in \cite{22}.
\item[3.] The authors of \cite{22} sketch the prospects that their model 
may have. The list of corresponding items is almost identical to that of 
\cite{19} which predates \cite{22}. Hence let us compare the models 
of \cite{22} and \cite{19}. The model of \cite{19} is based on 
the general model in \cite{0} rather than the special case of non rotational dust also discussed in \cite{0}. It thus performs, in contrast to 
\cite{22}, a complete reduction of 
the physical phase space since it contains also the scalar fields $S^j$.
The authors of \cite{19} outline a quantisation of the algebra 
of full quantum observables and the physical Hamiltonian $H$,  
which is isomorphic to the simple algebra of 
the $q,p$, on both Hilbert spaces ${\cal H}_{{\rm LQG}}$ and 
${\cal H}_{{\rm AQG}}$ which rightfully can be called the {\it physical 
Hilbert space} in contrast to the model \cite{22}. The price to pay is 
that the physical Hamiltonian is now given by the quantisation of the term 
\be \label{3.14}
H=\int\; d^3x\; \sqrt{|C^2-q^{ab} C_a C_b|}
\ee
The additional absolute value respects the fact that the quantity 
under the square root is constrained to be non negative in the classical
theory and in this
form (\ref{3.14}) allows for a well defined quantisation as discussed in 
\cite{19}.
Since the spatial diffeomorphism constraint reads 
\be \label{3.15}
C^{{\rm tot}}_a=P T_{,a}+P_j S^j_{,a}+C_a
\ee
even in the gauge $T=\tau$ and $S^j=\sigma^j$ (\ref{3.14}) does not reduce 
to $|C|[1]$. This is because in this gauge (\ref{3.15}) becomes
\be 
C^{{\rm tot}}_a=P_j \delta^j_{a}+C_a
\ee
However, this is neither a problem, since 
semiclassical tools are available in order to deal with square
root Hamiltonians \cite{23}, nor worse 
than working with $|C|$. 
\item[4.]
It is easy to see that the model 
\cite{22} corresponds to a subsector of the model \cite{19}. This can be 
demonstrated 
both at the classical and the quantum level. 
\\
\\
At the classical level
this corresponds to the observation that 
the portion of phase space where $C_a=0$ is left invariant
by (\ref{3.14}). Let us discuss the degrees of freedoms in both models. 
In \cite{22} one has gravity plus one scalar field and in \cite{19} gravity 
plus four scalar fields. In both models the total spatial diffeomorphism 
constraint is required to vanish yielding at the level of the gauge 
$T=\tau$ for the model \cite{22} the constraint $C_a=0$. When we consider 
for the model in \cite{19} the gauge $T=\tau$ and $S^j=\sigma^j$ we obtain 
$C_a+P_j\delta^{j}_a=0$ where $P_j$ are the
momenta conjugate to $S^j$. Let us define $P_a:=P_j\delta^{j}_a$ then in the 
model \cite{19} we have $C_a=-P_a$ and it can be shown that $P_a$ is a 
constant of motion and thus so is $C_a$. In particular, requiring $C_a$ 
to vanish in the model of $\cite{19}$ means $P_a=0$ yielding 3 additional 
constraints reducing the number of degrees of freedom to that of the 
model \cite{22}. Further evidence is obtained by noting that $P_a=0$ means 
$P_j=0$ and from this follows,
using $W_j\sim P_j$, that the Lagrange multiplier fields $W_j$ are 
constrained to vanish. This means that the reference fields $S^j$ are no 
longer present in the action. Hence we arrive at the special case of 
non-rotating dust, demonstrating the embedding at the classical level.
\\
\\
At the quantum level, the embedding is less intuitive but nevertheless can be shown.
Let us first discuss the case where we choose  the version of \cite{19}
where one quantises on ${\cal H}_{{\rm LQG}}$. One observes that the physical
Hamiltonian operator $H$ is invariant under diffeomorphism on the dust 
manifold $S(\sigma)$ and thus commutes with it generator $C_a$ where the index a here needs to be 
understood as labelling coordinates on the dust manifold $S(\sigma)$. One would rather call $C_a$ 
a generator of symmetries in dust space than a generator of gauge transformations.
This symmetry generator is implemented on the physical Hilbert space which is the standard kinematical Hilbert space
${\cal H}_{{\rm LQG}}$ of LQG. Now 
given the physical Hilbert space, one can look for functions that are invariant under the symmetry transformations
 generated by $C_a$ and those (dust)  diffeomorphism invariant 'functions', that are rather 
 distributions, will be the 'subset' of functions that are additionally annihilated by $C_a$ and hence
  also by the operator $q^{ab}C_aC_b$ when $q^{ab} C_a C_b$ is properly quantised as to 
annihilate diffeomorphism invariant distributions.
 On those diffeomorphism invariant distributions $H$ reduces to $|C|$. 
\\ 
Now, let us consider the second possibility where one chooses the version of 
\cite{19}
that is quantised on ${\cal H}_{{\rm AQG}}$ and see how the quantum theory 
of \cite{22} is embedded in that of \cite{19}.
Suppose we have fully reduced the Hamiltonian and spatial diffeomorphism 
constraint at the classical level, then the operator associated with the 
classical expression shown in (\ref{MAQG}) is a generator of 
symmetries rather than
 gauge transformations at the physical Hilbert space ${\cal H}_{{\rm AQG}}$. 
Likewise to the standard LQG quantisation, one can now look for 'functions' (distributions) that are annihilated by that operator and consider the solution space as a 
 'subspace' of the physical sector of the theory. However, here in general 
this 'subspace' will not be left invariant under the dynamics unless, 
as discussed before the operators $M,C$ (\ref{MAQG}) are quantised in such a 
way that they mutually commute. This would in any case be a desirable
feature of the theory but this is not granted by the naive quantisations
presented in \cite{23} and employed in \cite{22}. In other words, whenever
$M$ is defined in a satisfactory way in the model \cite{22} at the algebraic
level then it corresponds to a subsector to the algebraic version of 
\cite{19}. 
\\
Finally, we observe the following curiosity when working in the standard 
LQG framework. Let us consider the 'subspace' of functions 
that are invariant under the symmetry transformation generated by the 
$C_a$ on ${\cal H}_{{\rm LQG}}$. Now this 'subspace' has mathematically the 
same structure as the standard diffeomorphism invariant Hilbert space of 
LQG. Therefore, on this 'subspace' the physical Hamiltonian reduces to an 
expression where all terms involving $q^{ab}C_aC_b$ can be neglected and 
furthermore a graph-modifying quantisation could be applied. Hence, in this 
sense, in this subsector knot class modifying models can be rediscovered 
although it would be a bit artificial to force a knot class modifying 
quantisation in that particular 'subspace' whereas on the 'rest' of the 
Hilbert space a graph preserving quantisation is adopted.
\item[5.] A final comment concerns the claim made in \cite{22} that their
physical Hamiltonian is quantised free of ``anomalies'' as compared to 
\cite{23}.
Just to avoid confusion, let us try to
interpret this statement. If what is meant
that one single Hamiltonian operator (as compared to an infinite number
of Hamiltonian constraint operators) commutes with itself, the statement is 
empty. If what is meant is that the Hamiltonian densities (as compared to 
the Hamiltonian which is the integral of the densities) mutually commute
which is equivalent to the mutual commuting of the Hamiltonian constraint
operators, the statement is wrong at the algebraic level in which the 
authors of \cite{22} are working. The commutator of two Hamiltonian densities
is not an algebraic version of the diffeomorphism constraint and in that 
sense there is an anomaly. This is again the reason why in 
\cite{23} a master constraint approach towards all constraints has been 
adopted. If what is meant is that within the embedded LQG framework 
the Hamiltonian constraints commute on diffeomorphism invariant contributions,
this is correct when using the quantisation of \cite{28} but not when 
using the algebraic quantisation of \cite{23} that the authors copy. 
In any case the quantisation of \cite{28} is inappropriate for a physical 
Hamiltonian rather than an infinite number of Hamiltonian constraints because 
the resulting Hamiltonian would not even be symmetric in that quantisation
scheme (for constraints, symmetry is not necessary as one is only interested 
in the 
joint kernel). Rather, as pointed out in \cite{19}, one must use the graph 
preserving quantisation scheme (and a symmetric ordering) in order 
to obtain a symmetric operator which is spatially diffeomorphism 
invariant. Finally, as already pointed out in \cite{19}, the issue of 
anomalies is much less critical in this reduced phase space quantisation
approach as compared to the operator constraint quantisation approach
because the number of degrees of freedom has been correctly 
reduced already at the classical level. In that sense there can be no anomaly.
However, one might still be interested in implementing 
the classical hypersurface deformation algebra as a physical principle to 
reduce the quantisation ambiguities and this remains true for all models
considered so far.

\end{itemize}

\section{Gaussian Dust}
\label{s4}
It transpires that an optimal model would be such that 1. the physical 
Hilbert space is the usual ${\cal H}_{{\rm LQG}}$ (or ${\cal H}_{{\rm AQG}}$)
and 2. the physical Hamiltonian density is equivalent to just $C$ and not
$|C|$ or $\sqrt{|C^2-q^{ab} C_a C_b|}$. None of the models 
\cite{19,20,22} has both features.
Remarkably, we find exactly such 
a model in \cite{2}. In what follows we sketch the classical treatment
of this model in some detail since it differs slightly from that of 
\cite{15,16}. In contrast to \cite{2} we will not simply eliminate the Lagrange 
multiplier fields by their equations of motion but rather go through 
a careful Dirac treatment as outlined in section \ref{s2.5}. 
The quantisation of this model will be copied 
from \cite{19,28} and we will therefore be brief on that point.
For the physical interpretation of this model as a Gaussian reference fluid 
we refer to \cite{2}. We just mention here that variation of the action 
with respect to $\rho,W_j$ respectively leads to the conditions
$U^\mu U_\mu=-1,\;U_\mu V_j^\mu=0$ 
on the vectors $U^\mu=g^{\mu\nu} T_{,\nu},\;V_j^{\mu}:=g^{\mu\nu} S^j_{,\nu}$
and the equations of motion for $T,S^j$ yield the geodesic 
equation $\nabla_U U=0$ and the conservation equation 
$\nabla_\mu (W_j U^\mu)=0$. Hence the integral curves of $U$, labelled 
by $S^j(x)=\sigma_j=const.$ describe an 
observer in geodesic motion and the vectors $V_j$ are orthogonal to the 
corresponding $T=const.$ surfaces. Hence in the corresponding reference
frame the metric assumes the Gaussian form with unit lapse and zero shift.
The energy momentum tensor turns 
out to be $T^{\mu\nu}=\rho U^\mu U^\nu+\sum_j W_j V_j^{(\mu} U^{\nu)}$ 
which has no spatial trace with $U_\mu U_\nu+g_{\mu\nu}$ and thus no pressure
whence the name ``dust'' is appropriate. \\
\\
The Gau{\ss}ian dust action reads explicitly
\be \label{4.1}
{\cal L}_{{\rm GD}}=-
|\det(g)|^{1/2}\{\rho\frac{1}{2}[g^{\mu\nu} T_{,\mu} T_{,\nu}+1]
+g^{\mu\nu} T_{,\mu} [W_j S^j_{,\nu}]\}
\ee
Performing the 3+1 split in the ADM frame we obtain
\ba \label{4.2}
&&g^{\mu\nu} T_{,\mu} T_{,\nu}=-[L_n T]^2+q^{ab} T_{,a} T_{,b},\;\;
g^{\mu\nu} T_{,\mu} [W_j S^j_{,\nu}]=-[L_n T]\;[W_j L_n S^j]+
q^{ab} T_{,a} [W_j S^j_{,b}];\;\;
\nonumber\\
&& L_n=n=\frac{1}{N}[\partial_t-N^a\partial_a]
\ea
where $N,N^a$ are the usual lapse and shift functions and $q_{ab}$ is the 
intrinsic metric on the 3-manifold $\sigma$ (with inverse $q^{ab}$)
which is mapped via a one parameter
family of embeddings $Y_t$ into 
into a one parameter family of spacelike hypersurfaces $\Sigma_t=Y_t(\sigma)$
that foliate $M$. The timelike 
vector field $n$ is unit normal to the foliation and its action on the 
scalars in (\ref{4.2}) coincides with the Lie derivative $L_n$.

Performing the Legendre transform we find together with 
$\sqrt{|\det(g)}|=N\sqrt{\det(q)}$
\ba \label{4.3}
P &:=& \frac{{\cal L}_{{\rm GD}}}{\partial \dot{T}}=
\sqrt{\det(q)}\{\rho [L_n T]+W_j [L_n S^j]\}
\nonumber\\
P_j &:=& \frac{{\cal L}_{{\rm GD}}}{\partial \dot{S}^j}=
\sqrt{\det(q)}\;W_j\;[L_n T]
\nonumber\\
\pi &:=& \frac{{\cal L}_{{\rm GD}}}{\partial \dot{\rho}}=0
\nonumber\\
\pi^j &:=& \frac{{\cal L}_{{\rm GD}}}{\partial \dot{W}_j}=0
\ea
The detailed constraint analysis can be found in the appendix.
From it follows that we have 8 first class constraints $Z,\;Z_a,\;
C^{{\rm tot}},\;C^{{\rm tot}}_a$ and 8 second class constraints 
$z,z_j,\zeta_I,s,K$. Fortunately it is not necessary to compute 
the corresponding Dirac bracket explicitly by the following argument:\\
We arrange the second class constraints into 2 sets 
$\{K^{(1)}_\mu(x)\}_{I=1}^4=\{z,z^j\}(x)$ and 
$\{K^{(2)}_\mu(x)\}_{I=1}^4=\{\zeta_I,s,K\}(x)$. Then the Dirac matrix
\be \label{4.18}
\Delta^{IJ}_{\mu\nu}(x,y):=\{K^{(I)}_\mu(x),K^{(J)}_\nu(y)\}
\ee
and its inverse assumes a square block structure of the form
\be \label{4.19}
\Delta=\left( \begin{array}{cc} 
\Delta^{11}=0 & \Delta^{12} \\
-[\Delta^{12}]^T & \Delta^{22}
\end{array}
\right)\;\;
\Rightarrow\;\;
\Delta^{-1}=\left( \begin{array}{cc} 
([\Delta^{12}]^{-1})^T \Delta^{22} [\Delta^{12}]^{-1} & 
-([\Delta^{12}]^{-1})^T \\
{[}\Delta^{12}]^{-1} & 0
\end{array}
\right)
\ee
As we will eventually solve the second class constraints for 
the Lagrange multiplier fields $\pi,\pi^j,
\rho,W_j$ we are interested only in the restriction of the Dirac bracket
to functions $f$ of the variables $q,p;T,P;S^j,P_j$. For such functions 
we have $\{K^1_\mu(x),f\}=0$. Since the difference between 
the Dirac bracket and the Poisson bracket between $f,f'$
contains, due to (\ref{4.19}),
only terms with at least one of 
$\{K_\mu^1(x),f\}$ or $\{K_\mu^1(x),f'\}$, the Dirac bracket and the Poisson
bracket actually coincide on the functions of interest. 

The solution of the second class constraints $\zeta_I=K=s=0$  is given by
\ba \label{4.20}
W_I &=& W_3\frac{P_I}{P_3}
\nonumber\\
W_3^2 &=& \frac{P_3^2}{Q^2(1+q^{ab} T_{,a} T_{,b})}
\nonumber\\
\rho&=&
\frac{1}{P_3}\left(P-\frac{q^{ab} T_{,a} P_j S^j_{,b}}{1+q^{ab} T_{,a} T_{,b}}\right)
\ea
The solutions for $W_I,\rho$ are not explicitly needed but we need to choose 
a sign $\epsilon$ for $W_3/P_3$, insert the root of the second equation 
in (\ref{4.20}) into the Hamiltonian constraint
and see that the term proportional to $\rho$ vanishes identically
(since it enters linearly into the Hamiltonian constraint)
\be \label{4.21}
C^{{\rm tot}}=C+\epsilon\left(P\sqrt{1+q^{ab} T_{,a} T_{,b}}
+\frac{q^{ab} T_{,a} [P_j S^j_{,b}]}{\sqrt{1+q^{ab} T_{,a} T_{,b}}}\right)
\ee
while the spatial diffeomorphism constraint becomes
\be \label{4.22}
C^{{\rm tot}}_a=P T_{,a}+P_j S^j_{,a}+C_a
\ee
Substituting for $P_j S^j_{,a}$ in (\ref{4.21}) we find the equivalent
and simple form
\be \label{4.23}
C^{{\rm tot}}=C+\epsilon\frac{P-q^{ab} T_{,a} C_b}
{\sqrt{1+q^{ab} T_{,a} T_{,b}}}
\ee
or equivalently
\be \label{4.24}
C^{{\rm tot}}=P+h=:P+\epsilon C\sqrt{1+q^{ab} T_{,a} T_{,b}}-q^{ab}T_{,a}
C_b
\ee
We see that $h$ is of the type described in section (\ref{2.5}) so that 
the general theory outlined there applies. Moreover, we see a crucial 
difference with the model described in section \ref{s3}: The sign 
$\epsilon$ of $W_3/P_3$ is unrelated to the sign of $P,C$. 
The choice of $\epsilon$ in the Lagrangian also has no physical significance
since the $W_j$ dependent term is neither bounded from above nor from 
below. For definiteness we simply choose the phase space such that 
$\epsilon=+1$. Hence there is no absolute value of $C$ involved in
(\ref{4.24}) and the physical Hamiltonian 
becomes simply, following the exposition of section \ref{s2.5}
\be \label{4.25}
\hat{H}=\int_{{\cal S}}\;d^3\sigma\; \hat{C}(\sigma),\;
\hat{C}(\sigma)=C(\hat{q},\hat{p})
\ee
where ${\cal S}=S(\sigma)$ is the dust particle manifold and 
$\hat{q}=O'_{q'}(0),\;\hat{q}=O'_{p'}(0)$, that is, it is the same function
of $\hat{q},\hat{p}$ as is $C$ of $q,p$. Here $q',p'$ denotes the pull back
of the fields $q,p$ by the diffeomorphism $\sigma^j=S^j(x)$ 
and $O'_\cdot(0)$ is the projector map defined in section \ref{s2.5}. 
The Poisson algebras of the $q,p$ and of the $q',p'$ are identical.
Notice that (\ref{4.25}) is invariant under Diff$({\cal S})$.

The quantisation of the system can now be copied from \cite{28,19}. 
If we work on the Hilbert space ${\cal H}_{{\rm LQG}}$ and want the quantum 
operator to have the symmetries of its classical counterpart then 
$\hat{H}$ must preserve the subspaces ${\cal H}_{{\rm LQG},\gamma}$ 
defined by the closed linear span of SU(2) invariant spin network functions
over the graphs $\gamma$ embedded in $\cal S$ in order to be 
densely defined. In \cite{19} it is 
described how this can be achieved using the notion of a minimal loop
attachment and a corresponding projection operator onto 
${\cal H}_{{\rm LQG},\gamma}$. In the framework of ${\cal H}_{{\rm AQG}}$
the operator $\hat{H}$ does not need to preserve any of the subgraphs of 
the algebraic graph and can be defined in terms of next neighbour loops
\cite{19,23} just as in lattice gauge theory. In both approaches a symmetric
ordering of the Hamiltonian densities must be chosen. Here the following
subtlety arises:\\
When considering $|C|$ rather than $C$ the ordering of $C$ is 
chosen in such a way that $C$ acts only at the vertices of a graph.
This ordering is not symmetric but this does not matter because one 
considers the positive operator valued density $|C|:=\sqrt{C^\ast C}$. 
However, when quantising $C$ itself, such an ordering is not available.
To make sure that $C$ is symmetric and is densely defined (the danger being
that the curvature term involved in $C$ acts everywhere, not only at the 
vertices of a spin network function) one uses the tools developed in 
\cite{28} and writes 
\ba \label{4.26}
1&=&\frac{\left(\det(e(x))\right)^2}{\sqrt{\det(q(x))}^2}=
\left(\frac{\det(\{A(x),V_\epsilon(x)\})}{\sqrt{\det(q(x)}}\right)^2
\\
&=& \lim_{\epsilon\to 0}
\left(\frac{\det({\rm Tr}\left(A_\epsilon(x)^{-1}\{A_\epsilon(x),V_\epsilon(x)\}\right)}
{V_\epsilon(x)}\right)^2
\nonumber\\
&=& \left(\frac{3}{2}\right)^6 \lim_{\epsilon\to 0}
\left[\det\left(({\rm Tr}\left(A_\epsilon(x)^{-1}\{A_\epsilon(x),V_\epsilon(x)^{2/3}\}\right)\right)\right]^2
\nonumber
\ea
where $e(x)$ is the cotriad, $A(x)$ is the connection,
$V_\epsilon(x)$ is the volume function of a neighbourhood of $x$ 
with coordinate volume $\epsilon^3$ and $A_\epsilon(x)$ is a set of 
three SU(2) valued holonomy functions along three edges starting at $x$ 
with linearly independent tangents at $x$ and which span a coordinate volume
$\epsilon^3$ as well. As shown in \cite{28}, each of the two factors 
in (\ref{4.25}) admits a well defined quantisation
in the limit of vanishing regulator and acts only at vertices.
We can therefore freely position one factor each from (\ref{4.25}) to the 
outmost left and right of the operator and then order symmetrically (including
the projections on ${\cal H}_{{\rm LQG},\gamma}$ in the case of a quantisation 
on ${\cal H}_{{\rm LQG}}$).\\
\\
We notice that since $C$ is not a Kucha{\v r} density, the physical 
Hamiltonian has less symmetries than for the model described in 
\cite{15,19}, it does not Poisson commute with its density and thus 
it only has the Noether densities $\hat{C}_j(\sigma)$ as conserved charges.
However, $\hat{C}(\sigma)$ becomes a Noether charge when the other three
Noether charges $\hat{C}_j(\sigma)$ vanish because the $\hat{C}(\sigma),\;
\hat{C}_j(\sigma)$ obey the hypersurface deformation algebra. Thus, on this 
sector of the classical theory, the results of \cite{15,16,29} continue 
to hold which in the case of \cite{15,16} only assumed $\hat{C}(\sigma)$ 
to be preserved and non vanishing while $\hat{C}_j(\sigma)$ was kept 
arbitrary and in the case of \cite{29} one was working with 
$\hat{C}_j(\sigma)=0$ anyway. As a result, the model described in this section
is in agreement with the usual gravitational waves, cosmology, 
cosmological perturbation theory and black holes description as
as described by geodesic test observers, following the analysis  
of \cite{15,16,19}.

\section{Summary and Outlook}
\label{s5}

In this paper we have accomplished three tasks:
\begin{itemize}
\item[1.] We have described all matter models considered so far for the 
purpose of deparametrisation of General Relativity with an eye towards
Quantum Gravity starting from a general Lagrangian. This serves to bring 
order into the space of models studied already and those that can be within
the theory space described. We have described the Hamiltonian analysis 
of these models in a uniform fashion.
\item[2.] Basically two types of models have been constructed: II. Those 
which deparametrise only time and I. those which deparametrise spacetime. 

We have shown that those of type II. necessarily must implement spatial 
diffeomorphism invariance in the quantum theory. This is a non trivial 
subject for several reasons:\\ 
First, even though the unitary implementation of spatial diffeomorphisms
in LQG is no problem, there is no unique Hilbert space structure on the 
diffeomorphism invariant distributions \cite{27}. To fix the ambiguities, 
the implementation of the $^\ast$relations among spatially diffeomorphism 
invariant observables may be consulted as a guiding principle, however,
neither is a complete and algebraically independent set of such objects 
known even at the classical level nor is it clear that these 
can be quantised as to yield an honest representation of their 
$^\ast$algebra (see \cite{30} for the tremendous 
difficulties encountered already in the much simpler setting of the closed 
bosonic string). This is due to the fact that spatially diffeomorphism 
invariant functions on phase space are typically not simple polynomials of 
the elementary fields. The considerations made in \cite{21} indicate that
such anomalies are a conceivable possibility and that no choice of the 
ambiguity parameters may exist for which the representation problem can
be solved. 

These difficulties are naturally avoided in the type I models where 
also the spatial diffeomorphism symmetry is reduced already at the 
classical level. The non trivial observation is that this can be done 
while keeping a very simple Poisson $^\ast$algebra of basic fields so 
that one can easily find Hilbert space representations thereof. This 
also extends to the reduction of the symmetries generated by the Hamiltonian 
constraint, at least for the matter models considered here. This 
is quite remarkable when recalling the difficulties that one meets when 
trying to implement the Hamiltonian operator constraint in LQG, see 
e.g. the discussion in \cite{19,23,25,31}.
\item[3.] The authors of \cite{22} point out that their model, which belongs
to type II, avoids a square root in the physical Hamiltonian density in 
contrast 
to the models considered in \cite{19,20}. We have shown that this 
statement has to be made more precise. If one insists on the usuall 
energy conditions of classical General Relativity
then the analysis shows that rather $|C|=\sqrt{C^2}$ is the 
physical Hamiltonian density which again needs a square root. We remark that 
square roots are a nuisance rather than a caveat as semiclassical tools 
are available that can deal with this problem \cite{23}. However, it is 
certainly true that the analysis simplifies significantly when there 
is no square root. Unfortunately, in type II models the quantum reduction 
of the spatial diffeomorphism constraint still poses a challenge.
In view of what was just said, this leads to the natural question 
whether there are matter models within type I which lead to a physical 
Hamiltonian without square root. It is a remarkable achievement of Kucha{\v r}
and Torre \cite{2} to have achieved just that in the form of the 
Gaussian dust. We have applied essential steps of the analysis 
performed in \cite{15} to the model \cite{2} in order to show that all that 
was said in \cite{15,16,19,29} continues to be correct. We believe that 
in the class of models of type II considered, the degree of simplicity 
of the physical Hamiltonian cannot be improved over the model presented 
in section \ref{s4}.
\end{itemize}
We believe that switching from the earlier operator constraint reduction 
(Dirac) approach that dominated the research in LQG over the past 20 
years to the reduced phase space approach, concretely implemented 
within LQG for the first time in \cite{15,19} has many merits. 
Within the former approach, there are many steps of significant 
technical complexity to be overcome before one reaches the same level 
of technical and conceptual control as in the latter. After all, ultimately 
also in the former approach one aims at {\it the physical inner product,
physical vector states, physical observables and the physical dynamics} which 
come for free in the latter approach. Moreover, on physical grounds the 
intrinsic description of the physics in terms of material reference 
systems makes a lot of sense, hence the approach advocated here and by 
many others before is well motivated. 

There are certainly some possible 
caveats: For instance, the classical reduction assumes that certain 
objects such as $\det(\partial S/\partial x)$ is everywhere non vanishing.
As one can show, this is a gauge invariant condition but it imposes a 
non holonomic constraint on the phase space. Hence, the description is 
not valid in the full unconstrained phase space. On the other hand, 
these issues are maybe less important than one thinks as the material
reference system serves a role quite similar to three of the four 
real degrees of freedom of the Higgs field which give mass to $W^\pm$ and $Z$ 
bosons: They are simply absorbed into those vector bosons thus making them
local Isospin SU(2) singlets. Similarly here, the dust field is absorbed into 
the geometry and matter degrees of freedom, thus making them local Diff$(M)$ 
singlets. Hence the material reference system is the gravitational counterpart
of the Higgs effect.\\
\\
The real physical question is whether scalar fields such as the Higgs field
or the dust considered here exist in nature. The answer might quite 
well be negative. Although the dust field considered here and elsewhere
in some sense make it an ideal dark matter candidate since it is only 
interacting with gravity and with itself, it may nevertheless be 
phenomenologically excluded. What we need is maybe
a realistic dark matter candidate, to be found in some extension of the 
standard model, that can serve the purpose of deparametrisation. 
It would be perhaps most economic to isolate four degrees of freedom from 
the spacetime metric tensor for the purposes of deparametrisation but 
this is more difficult than with matter because one needs to construct 
scalars using covariant derivatives which makes such a construction non local
and thus impractical.\\
\\ 
In any case, dust fields and their generalisations as described in the 
present work provide a proof of principle that a reduced phase space 
quantisation
approach to LQG not only works but moreover has many advantages 
over the Dirac operator constraint approach  
because it lifts all the structures 
previously found in LQG directly to the physical versus kinematical level.
In view of these advantages, the other caveats mentioned are in our mind
problems of lower priority.\\
\\
\\
\\
{\bf\large Acknowledgements}\\
K.G. would like to thank Jerzy Lewandowski for illuminating discussions.
\\
\\
\begin{appendix}
\section{Constraint Analysis for the  Gaussian Dust Model}
In this section the constraint analysis of the Gaussian dust model is presented.
From the Legendre transform 
we deduce the following primary constraints
\be \label{4.4}
z=\pi,\;z^j=\pi^j,\;Z=\Pi,\;Z_a=\Pi_a,\;R^j:=\epsilon^{jkl} P_k W_l
\ee
with the momenta $\Pi=\Pi_a=0$ for 
lapse and shift functions respectively. We notice that only two of the 
3 constraints $R^j$ are linearly independent because obviously 
$W_j R^j=0$. In what follows we will assume that 
$\sum_j W_j^2>0$ and $\sum_j P_j^2>0$. Then in a patch with $W_3,P_3\not=0$ 
we can choose
\be \label{4.5}
\zeta_1:=P_1-\frac{W_1}{W_3} P_3=-R^2/W_3,\;
\zeta_2:=P_2-\frac{W_2}{W_3} P_3=R^1/W_3
\ee
as the independent ones which tell that all $P_j$ are proportional to $P_3$.
The velocity combinations that
can be solved for in (\ref{4.3}) are
\be \label{4.6}
[L_n T]=\frac{P_3}{W_3 \sqrt{\det(q)}},\;\;
W_j [L_n S^j]=\frac{P}{\sqrt{\det(q)}}-\frac{\rho P_3}{W_3 \sqrt{\det(q)}}
\ee
We compute the canonical Hamiltonian density using the abbreviation 
$Q=\sqrt{\det(q)}$
\ba \label{4.7}
{\cal H}_{{\rm GD,can}}
&=&
[P\dot{T}+P_j\dot{S}^j+\pi\dot{\rho}+\pi^j\dot{W}_j+\Pi\dot{N}+\Pi_a\dot{N}^a
-{\cal L}_{{\rm GD}}]_{{\rm (\ref{4.6})}}
\nonumber\\
&=& 
\{N^a[P T_,a+P_j S^j_{,a}]+uz+u_j z^j+VZ+V^a Z_a+v^I \zeta_I
+N(P [L_n T]+\frac{P_3}{W_3} [W_j L_n S^j])
\nonumber\\
&& +N Q\{\frac{1}{2}\rho[-[Ln T]^2+q^{ab} T_{,a} T_{,b}+1]
+[-[L_n T][W_j [L_n S^j]]+q^{ab} T_{,a} [W_j S^j_{,b}]\}_{{\rm (\ref{4.6})}}
\nonumber\\
&=& 
N^a[P T_,a+P_j S^j_{,a}]+uz+u_j z^j+VZ+V^a Z_a+v^I \zeta_I
+N(P \frac{P_3}{W_3 Q}+\frac{P_3}{W_3} [\frac{P}{Q}-\frac{P_3 \rho}{W_3 Q}])
\nonumber\\
&& +NQ\{\frac{1}{2}\rho[-[\frac{P_3}{W_3 Q}]^2+q^{ab} T_{,a} T_{,b}+1]
+[-[\frac{P_3}{W_3 Q}][\frac{P}{Q}-\frac{P_3 \rho}{W_3 Q}]
+q^{ab} T_{,a} [W_j S^j_{,b}]\}
\nonumber\\
&=& 
N^a[P T_,a+P_j S^j_{,a}]+uz+u_j z^j+VZ+V^a Z_a+v^I \zeta_I
+N\{
\frac{P P_3}{W_3 Q}-\frac{\rho}{2 Q}[\frac{P_3}{W_3}]^2\}
\nonumber\\
&& +NQ\{\frac{1}{2}\rho[q^{ab} T_{,a} T_{,b}+1]+q^{ab} T_{,a} [W_j S^j_{,b}]\}
\nonumber\\
&=& 
N^a[P T_,a+P_j S^j_{,a}]+uz+u_j z^j+VZ+V^a Z_a+
[v^I-N\frac{W_3}{P_3}Q q^{ab} T_{,a} S^I_{,b}]  \zeta_I
\nonumber\\
&& +N\{
\frac{P P_3}{W_3 Q}-\frac{\rho}{2 Q}[\frac{P_3}{W_3}]^2\}
\nonumber\\
&& +NQ\{\frac{1}{2}\rho[q^{ab} T_{,a} T_{,b}+1]+
\frac{W_3}{P_3} q^{ab} T_{,a} [P_j S^j_{,b}]\}
\ea
where we used the abbreviations 
\be \label{4.7a}
u=\dot{\rho},\;u_j=\dot{W}_j,\;V=\dot{N},\;V^a=\dot{N}^a,\;
v^I=L_n S^I;\;I=1,2
\ee
The contributions to the canonical Hamiltonian from geometry and standard 
matter are $N^a C_a+N C$ which has to be added to (\ref{4.7}) in order 
to obtain the total canonical Hamiltonian 
\be \label{4.8}
H_{{\rm can}}=\int_\sigma\;d^3x \; ({\cal H}_{{\rm GD,can}}+NC+N^a C_a)
\ee
Stability of the primary constraints yields on the constraint surface 
of the primary constraints
\ba \label{4.9}
C^{{\rm tot}} &:=& \{Z,H_{{\rm can}}\}=C+
\frac{1}{Q}[\frac{P P_3}{W_3}-\frac{\rho}{2}[\frac{P_3}{W_3}]^2]
+Q[\frac{1}{2}\rho[q^{ab} T_{,a} T_{,b}+1]+
\frac{W_3}{P_3}q^{ab} T_{,a} [P_j S^j_{,b}]
\nonumber\\
C^{{\rm tot}}_a &:=& \{Z_a,H_{{\rm can}}\}=C_a+P T_{,a}+P_j S^j_{,a}
\nonumber\\
\frac{N}{2} s &:=& \{z,H_{{\rm can}}\}=
\frac{N}{2}\{-\frac{1}{Q}[\frac{P_3}{W_3}]^2
+Q[q^{ab} T_{,a} T_{,b}+1]\}
\nonumber\\
&& \{z_I,H_{{\rm can}}\}=
-[v^I-N\frac{W_3}{P_3}Q q^{ab} T_{,a} S^I_{,b}]\frac{P_3}{W_3}
\nonumber\\
N K &:=& \{z_3,H_{{\rm can}}\}=
[v^I-N\frac{W_3}{P_3}Q q^{ab} T_{,a} S^I_{,b}]\frac{W_I P_3}{W_3^2}
+N\{-\frac{P P_3}{W_3^2 Q}+\frac{\rho}{Q}\frac{P_3^2}{W_3^3}
+Q\frac{1}{P_3} q^{ab} T_{,a} [P_j S^j_{,b}]\}
\nonumber\\
&& \{\zeta_I,H_{{\rm can}}\}=u_I\frac{P_3}{W_3}+M_I
\ea
where we used that $\{\zeta_I,\zeta_J\}=0$ and $M_I$ is some complicated 
and non vanishing expression whose explicit form will be of no further
interest. The set of equations (\ref{4.9}) has to vanish which is accomplished
by choosing
\be \label{4.10}
v^I=N\frac{W_3}{P_3}Q q^{ab} T_{,a} S^I_{,b},\;
u_I=-\frac{P_3}{W_3}M_I
\ee
and imposing the secondary constraints $C^{{\rm tot}},\;
C^{{\rm tot}}_a,\; s,\;K$. Notice that due to (\ref{4.10}) $K$ simplifies 
to 
\be \label{4.11}
K=-\frac{P P_3}{W_3^2 Q}+\frac{\rho}{Q}\frac{P^2_3}{W_3^3}
+Q\frac{1}{P_3} q^{ab} T_{,a} [P_j S^j_{,b}]
\ee
and the terms proportional to $\zeta_I$ drop from the Hamiltonian.

Turning to the stabilisation of the secondary constraints, we add to 
$C^{{\rm tot}}_a$ the linear combination of the already stabilised
primary constraints $\pi\rho_{,a}+\pi^j W_{j,a}+\Pi N_{,a}+
\Pi_b N^b_{,a}+(\Pi_a N^b)_{,b}$. Then $C^{{\rm tot}}_a$
generates spatial diffeomorphisms on all variables involved and since 
${\cal H}_{{\rm can}}$ is a spatial scalar density of weight one, 
$C_a^{{\rm tot}}$ is stabilised.   
Next we have on the constraint surface determined so far
\ba \label{4.12}
0&=& \{C^{{\rm tot}},H_{{\rm can}}\}=
\{C^{{\rm tot}},C^{{\rm tot}}[N]\}
-u\frac{\partial C^{{\rm tot}}}{\partial \rho}
-u^3\frac{\partial C^{{\rm tot}}}{\partial W_3}
\nonumber\\
&=& 
\{C^{{\rm tot}},C^{{\rm tot}}[N]\}-\frac{u}{2}s -u^3 K
\nonumber\\
&=& 
\{C^{{\rm tot}},C^{{\rm tot}}[N]\}
\nonumber\\
0 &=& \{s,H_{{\rm can}}\}
=-2u^3\frac{1}{Q}\frac{P^2_3}{W_3^3}+M
\nonumber\\
0&=& \{K,H_{{\rm can}}\}
=-u\frac{1}{Q}\frac{P^2_3}{W_3^3}+M'
\ea
where $M$ is independent of $u$ and $M'$ depends linearly on $u^3$. 
We can therefore solve the two last equations for $u,u^3$ respectively
so that $s,K$ are stabilised. As far as the first term is concerned 
we write for some smearing function $f$ 
\ba \label{4.13}
C_{{\rm GD}}[f] &:=& \int \; d^3x \; f\;(T+U)
\\
&:=& \int \; d^3x \; f\;
\{\frac{1}{Q}(
\frac{P P_3}{W_3}-\frac{\rho}{2}[\frac{P_3}{W_3}]^2)
+Q(\frac{1}{2}\rho[q^{ab} T_{,a} T_{,b}+1]+
\frac{W_3}{P_3} q^{ab} T_{,a} [P_j S^j_{,b}])\}
\nonumber
\ea
and similar for $C[f]$. Then 
\be \label{4.14}
\{C^{{\rm tot}}[f],C^{{\rm tot}}[f']\}
=\{C[f],C[f']\}
+\{C[f],C_{{\rm GD}}[f']\}
-\{C[f'],C_{{\rm GD}}[f]\}
+\{C_{{\rm GD}}[f],C_{{\rm GD}}[f']\}
\ee
The first term gives $-C_a[q^{ab}(f f'_{,b}-f_{,b} f')$ as is well known 
from the hypersurface deformation algebra \cite{26}. The second and 
first term cancel each other because the only piece from $C$
that contributes is the gravitational piece which acts ultralocally only 
on the $q_{ab}$ dependence in $C_{{\rm GD}}$ which thus gives due 
to the non derivative coupling a term proportional to 
$\delta(x,y)[f(x)f'(y)-f(y) f'(x)]=0$. The last term is of a new type.  
Again due to vanishing ultralocal terms we just need to focus on terms 
that lead to derivatives of the $\delta$ distributions. For the same 
reason we only need to keep track of the $x,y$ dependence of the smearing
fields. Accordingly in the following calculation we neglect ultralocal 
contributions 
\ba \label{4.15}
\{C_{{\rm GD}}[f],C_{{\rm GD}}[f']\}
&=&\frac{1}{2}\int\; d^3x \int d^3y [f(x) f'(y)-f(y) f'(x)]
\{C_{{\rm GD}}(x),C_{{\rm GD}}(y)\}
\nonumber\\
&=&\frac{1}{2}\int\; d^3x \int d^3y [f(x) f'(y)-f(y) f'(x)]
\nonumber\\
&& \times [\{T(x),T(y)\}+\{T(x),V(y)\}-\{T(y),V(x)\}+\{V(x),V(y)\}]
\nonumber\\
\{T(x),T(y)\} &=&=0
\nonumber\\
\{T(x),V(y)\} &=& 
\frac{1}{W_3}\{(PP_3)(x),
(\frac{1}{2}\rho[q^{ab} T_{,a} T_{,b}+1]+
\frac{W_3}{P_3} q^{ab} T_{,a} [P_j S^j_{,b}])(y)\}
\nonumber\\
&& - 
\frac{\rho P_3}{W_3^2}\{ P_3(x),
(\frac{1}{2}\rho[q^{ab} T_{,a} T_{,b}+1]+
\frac{W_3}{P_3} q^{ab} T_{,a} [P_j S^j_{,b}])(y)\}
\nonumber\\
&=&
\frac{P_3}{W_3} q^{ab}[\rho T_{,a}+\frac{W_3}{P_3} P_j S^j_{,a}]
+\frac{P}{W_3} q^{ab} T_{,a} \frac{W_3}{P_3} P_j\delta^j_3  
\delta_{,y^b}
\nonumber\\
&& -\frac{\rho P_3}{W_3^2} \frac{W_3}{P_3} q^{ab} T_{,a} P_j \delta^j_3
\delta_{,y^b}
\nonumber\\
&=& \delta_{,y^b} q^{ab}[P_j S^j_{,a}+P T_{,a}]
\nonumber\\
\{V(x), V(y)\} 
&=& Q^2 q^{ab} T_{,b} q^{cd} T_{,d} W_3^2
\{(P_j S^j_{,b}/P_3)(x),(P_k S^k_{,d}/P_3)(y)\}
\nonumber\\
&=& Q^2 q^{ab} T_{,b} q^{cd} T_{,d} W_3^2
S^j_{,b}
\{\frac{P_j}{P_3}(x),S^k_{,d}(y)\} \frac{P_k}{P_3} -x\leftrightarrow y
\nonumber\\
&=& Q^2 q^{ab} T_{,b} q^{cd} T_{,d} W_3^2
S^I_{,b}
\{\frac{P_I}{P_3}(x),S^k_{,d}(y)\} \frac{P_k}{P_3} -x\leftrightarrow y
\nonumber\\
&=& Q^2 q^{ab} T_{,b} q^{cd} T_{,d} W_3^2
S^I_{,b}
[\{P_I(x),S^k_{,d}(y)\} \frac{P_k}{P^2_3} 
-\{P_3(x),S^k_{,d}(y)\} \frac{P_k P_I}{P_3^3}]
-x\leftrightarrow y
\nonumber\\
&=& Q^2 q^{ab} T_{,b} q^{cd} T_{,d} W_3^2
S^I_{,b} \delta_{,y^d}
[\frac{P_I}{P^2_3} 
-\frac{P_3 P_I}{P_3^3}]
-x\leftrightarrow y
\nonumber\\
&=& 0
\ea
It follows 
\ba \label{4.16}
\{C_{{\rm GD}}[f],C_{{\rm GD}}[f']\}
&=&\int\; d^3x \int d^3y [f(x) f'(y)-f(y) f'(x)]
\{T(x),V(y)\}
\nonumber\\
&=& \int\; d^3x \int d^3y [f(x) f'(y)-f(y) f'(x)]
\delta_{,y^b} q^{ab} C_{{\rm GD}a}
\nonumber\\
&=& -C_{{\rm GD}}[q^{ab}(f f'_{,b} -f_{,b} f')]  
\ea
and therefore 
\be \label{4.17}
\{C^{{\rm tot}}[f],C^{{\rm tot}}[f']\}
= -C_a^{{\rm tot}}[q^{ab}(f f'_{,b} -f_{,b} f')]  
\ee
satisfies the hypersurface deformation algebra. All constraints are 
now stabilised. 

The constraints $Z,Z_a$ are trivially first class. The constraint 
$C^{{\rm tot}}_a$ is first class as we have already seen due to its 
geometrical action. The constraints $z_I,\zeta_I,\;I=1,2$ form second class 
partners as well as $(z,K)$ and $(z_3,s)$.
Finally, the constraint $C^{{\rm tot}}$ closes to itself with 
$Z,Z_a,C^{{\rm tot}}_a$ and with $z,z_3$ respectively it closes to
$s,K$ respectively. To make it close with $z_I,\zeta_I,s,K$ as well
we add suitable linear combinations of all second class constraints
to $C^{{\rm tot}}$ so that it has vanishing Poisson brackets with all 
second class constraints. 
\end{appendix}

\end{document}